\documentclass[preprint,review]{article}
\usepackage[german, english]{babel}
\usepackage[a4paper, left=2.5cm, right=2.5cm, top=3.5cm, bottom=3cm]{geometry}
\usepackage{amssymb}
\usepackage{amsmath}
\usepackage{amscd}
\usepackage{latexsym}
\usepackage{graphicx}
\usepackage{csquotes}
\usepackage{ifthen}
\usepackage{epsfig}
\usepackage{setspace}
\usepackage{hyperref}
\usepackage{dsfont}
\usepackage{mathrsfs}
\usepackage{color}
\usepackage{cancel}
\usepackage{mathtools}
\usepackage{wrapfig}
\usepackage{float}
\usepackage{subfig}
\hypersetup{
 colorlinks=true,
 linkcolor=blue,
 citecolor=magenta,
}

\catcode`@=11
\renewcommand{\section}{\@startsection{section}{1}{0pt}{\medskipamount}
{\medskipamount}{\large\bf}}
\numberwithin{equation}{section}
\catcode`@=12

\newcommand{\Z}{\mathbb{Z}}
\newcommand{\R}{\mathbb{R}}

\newcommand{\Fcal}{{\cal F}}

\def\im{\mathrm{i}}
\def\ep{\mathrm{e}}
\def\pa{\partial}
\def\diff{\mathrm{d}}
\def\tr{\mathrm{tr}}
\def\sfrac#1#2{{\textstyle\frac{#1}{#2}}}
\def\]{\right]}
\def\[{\left[}
\def\){\right)}
\def\({\left(}
\def\>{\rangle}
\def\<{\langle}
\def\+{\dagger}
\def\we{{\wedge}}
\def\={\ =\ }
\def\und{\quad\textrm{and}\quad}
\def\with{\quad\textrm{with}\quad}

\begin{document}

\title{\bf\huge 
Exact gauge fields from anti-de Sitter space}
\date{~}

\author{\phantom{.}\\[12pt]
\hspace{-.8cm}
{\Large Savan Hirpara}, \ 
{\Large Kaushlendra Kumar}, \
{\Large Olaf Lechtenfeld}, \
{\Large Gabriel Pican{\c c}o Costa} \
\\[24pt]
{Institut f\"ur Theoretische Physik \&}\\ 
{Riemann Center for Geometry and Physics}\\
{Leibniz Universit\"at Hannover} \\ 
{Appelstra{\ss}e 2, 30167 Hannover, Germany}\\
[12pt] 
} 
\maketitle

\begin{abstract}
\noindent\large
In 1977 L\"uscher found a class of SO(4)-symmetric SU(2) Yang--Mills solutions in Minkowski space,
which have been rederived 40 years later by employing the isometry $S^3\cong\mathrm{SU}(2)$ and 
conformally mapping SU(2)-equivariant solutions of the Yang--Mills equations on (two copies of)
de Sitter space $\mathrm{dS}_4\cong\mathbb{R}{\times}S^3$.
Here we present the noncompact analog of this construction via $\mathrm{AdS}_3\cong\mathrm{SU}(1,1)$. 
On (two copies of) anti-de Sitter space $\mathrm{AdS}_4\cong\mathbb{R}{\times}\mathrm{AdS}_3$ 
we write down SU(1,1)-equivariant Yang--Mills solutions and conformally map them to $\mathbb{R}^{1,3}$.
This yields a two-parameter family of exact SU(1,1) Yang--Mills solutions on Minkowski space,
whose field strengths are essentially rational functions of Cartesian coordinates.
Gluing the two AdS copies happens on a $\mathrm{dS}_3$ hyperboloid in Minkowski space, and our Yang--Mills 
configurations are singular on a two-dimensional hyperboloid $\mathrm{dS}_3\cap\mathbb{R}^{1,2}$.
This renders their action and the energy infinite, although the field strengths fall off fast 
asymptotically except along the lightcone.
We also construct Abelian solutions, which share these properties but are less symmetric and of zero action.
\end{abstract}

\thispagestyle{empty}
\newpage
% \clearpage

\section{Introduction and summary}
\setcounter{page}{1} 
\noindent
Analytic solutions to the Yang--Mills equations are hard to come by, especially in the absence of matter (Higgs) fields (for some reviews, see~\cite{actor,rajaraman,manton}).
Since pure Yang--Mills theory is conformally invariant in four spacetime dimensions, a classical field configuration on a suitable spacetime may be carried over to a conformally related background by means of a conformal transformation. In particular, a solution of the four-dimensional Yang--Mills equations on a conformally flat manifold provides us (at least locally) with an exact Yang--Mills field on Minkowski space and, more generally, on any Friedmann--Lema\^itre--Robertson--Walker universe. This idea has been used in~\cite{zhilin,kumar} for SU(2) and U(1) gauge theory on de Sitter space, to reproduce a Yang--Mills solution known since 1977~\cite{luescher} and to generate a new basis for electromagnetic knot solutions~\cite{ranada,knots}, respectively.

The success of this method relies on an identification of the gauge group with (a subgroup of) the isometry of the leaves of a spacetime foliation, which admits a symmetric (``equivariant'') ansatz. In~\cite{IvLePo1,IvLePo2} this was exercised to derive finite-action SU(2) Yang--Mills fields on $S^3$-foliated dS$_4$ and also, with restricted success, on AdS$_4$. Furthermore, the generalization to SO(4) solutions and to higher-dimensional de~Sitter spaces was performed in~\cite{uenal}.

While the cases mentioned above employ foliations of spacetime with compact submanifolds and thus compact gauge groups, the extension to the noncompact case is obvious geometrically. Indeed, for noncompact $H^3$ and dS$_3$ foliations of (parts of) Minkowski space, exact SO(1,3) Yang--Mills solutions were constructed recently~\cite{roehrig}. In this paper, we analyze the noncompact variant of the previously mentioned SU(2) gauge theory on dS$_4$, applying the aforestated construction method to SU(1,1) Yang--Mills on AdS$_3$-foliated anti-de Sitter space~AdS$_4$. While the AdS$_3$ leaves admit an SO(2,2) action, we follow the dS$_4$ example and choose the gauge group to be SL$(2,\R)\cong\textrm{SU}(1,1)$, which can be directly identified with the AdS$_3$ slices, providing an orthonormal frame of one-forms $\{e^0,e^1,e^2\}$ on AdS$_3$. In this way we obtain a two-parameter family of SU(1,1)-equivariant Yang--Mills fields on AdS$_4$ (with some radius~$R$), which depend in a particular way on a spatial foliation parameter~$\psi\in(-\frac{\pi}{2},\frac{\pi}{2})\equiv\mathcal{I}_\psi$.

Even though we are going to map the solutions to Minkowski space, one may question the physical plausibility of a gauge theory in AdS$_4$, due to the presence of closed timelike curves and the lack of global hyperbolicity. However, those issues can be effectively addressed. To tackle the former, one can ``unwrap" the periodic time coordinate and work in the universal covering space $\widetilde{\textrm{AdS}}_4$. Addressing the latter, both spaces feature a timelike boundary at spatial infinity, which poses complications for the Cauchy problem. Different solutions to this issue were discussed in~\cite{AIS78} utilizing special boundary conditions and constructing an ``effective" Cauchy surface in AdS$_4$. These questions would be especially serious for the quantum theory, as would be the fact that our gauge group is non-compact\footnote{
See, however, the discussion of unitary $S$-matrices for non-compact gauge theories in~\cite{margolin}.}.
Therefore, in this work we restrain ourselves to the classical level.

Conformally mapping $\mathcal{I}_\psi{\times}\textrm{AdS}_3\ni\{\tau,\rho,\psi,\phi\}$ to Minkowski space $\R^{1,3}\ni\{t,x,y,z\}$ is more complicated than for the compact case of $\mathcal{I}_\tau{\times}S^3$, because (a) the foliation parameter~$\psi$ is spatial and (b) only a circle $S^1\ni\{\phi\}$ is mapped isometrically rather than a two-sphere. For this reason, we first express the Yang--Mills solution in terms of intermediate $H^3$-slicing coordinates $\{\tau,\lambda,\theta,\phi\}\in S^1_\tau{\times}H^3$ with a temporal foliation parameter~$\tau\in S^1_\tau$, whose leaves are isomorphic to a hemisphere $S^3_+$ upon reparametrizing $\lambda\to\chi$ so that $\{\chi,\theta,\phi\}\in S^3$, and then employ the known Carter--Penrose map for half of dS$_4$ to push forward the solution to parts of Minkowski space. The conformal AdS$_4$ boundary isomorphic to $S^1_\tau{\times}S^2\ni\{\tau,\lambda{=}\infty,\theta,\phi\}$ (corresponding to the $S^3$ equator's $\chi{=}\frac{\pi}{2}$ locus) is mapped to the one-sheeted hyperboloid $x^2{+}y^2{+}z^2{-}t^2{=}R^2$, and the domain $S^1_\tau{\times}H^3\cong S^1_\tau{\times}S^3_+$ (corresponding to $\chi\in[0,\frac{\pi}{2})$) is mapped to its exterior. In order to cover the entire Minkowski space, we need to glue on a second AdS$_4$ copy, whose leaves are taken to form the other hemisphere~$S^3_-$ (corresponding to $\chi\in(\frac{\pi}{2},\pi]$). Comparing to the previous construction via de Sitter space~\cite{zhilin,kumar} where two copies of~dS$_4$ could be glued ``on top of each other'' smoothly, the AdS case considered here requires a ``sideways'' gluing of two copies of~AdS$_4$, which in contrast produces a singularity.

In this way we arrive at a family of exact SU(1,1) Yang--Mills solutions on Minkowski spacetime, whose Riemann--Silberstein vector is essentially a rational function. Their energy-momentum tensor is found to be singular at the intersection of the boundary hyperboloid with the hyperplane $\{z{=}0\}$, which is a two-dimensional Lorentzian hyperboloid. Unfortunately, this singularity is not integrable and thus leads to a divergent total energy and action of the field configuration.
At the same time, a (non-equivariant) restriction to Abelian field configurations is always possible and yields harmonic equations for the ansatz function (of the foliation parameter). We provide the electric and magnetic field strengths for the Maxwell solution obtained by restricting $\textrm{SU}(1,1)\to\textrm{U}(1)$. This not only provides a ``hyperbolic'' version of the Ra\~nada--Hopf electromagnetic knot~\cite{ranada,knots} but also reproduces, as a special case, a recently found magnetic vortex configuration on $\R^{1,2}$~\cite{RS18}. Its energy density is again non-integrable while the action vanishes. 
Other variants of our Maxwell solution, where the Abelian subgroup is embedded differently into SU(1,1), remain to be interpreted physically.
The physical codimension-2 singularity may be sourced by adding appropriate matter to the system, such as a circular 1-brane.

Besides the two dimensionless family parameters, the Yang--Mills and Maxwell solutions constructed here depend on a single scale, the AdS radius~$R$, which provides the canonical dimensionality of all quantities. Since the map to Minkowski space selects a particular spatial direction (which we took to be the $z$~direction), the field configurations depend on~$x^2{+}y^2$ and $z$ separately. 
Our (color-)electromagnetic fields decay with the inverse fourth power of the (spatiotemporal) distance from the coordinate origin, except with the inverse first power along the lightcone. Seen from afar they resemble instantonic events.

%~~~~~~~~~~~~~~~~~~~~~~~~~~~~~~~~~~~~~~~~~~~~~~~~~~~

% \newpage

\section{Geometry of anti-de Sitter space $\text{AdS}_{4}$}

\noindent
$\mathrm{AdS}_{4} \cong {O(2,3)/O(1,3)}$ space is a hypersurface in $\mathbb{R}^{2,3}$, isometrically embedded into $\mathbb{R}^{2,3}$ via the relation
\begin{align}
    -(x^{1})^2-(x^{2})^2+(x^{3})^2+(x^{4})^2+(x^{5})^2\=-R^2\ .\label{2.1} 
\end{align}
We are interested in foliations $\mathrm{AdS_{4}}\cong{\cal I}\times \mathcal{M}_{3}$, where $\cal I$ is a real interval, and $\mathcal{M}_{3}$ is a three-dimensional homogeneous space. Two examples of such embedding are presented here:

\noindent
\textbf{AdS$_3$-slicing coordinates.} 
In this case $\mathcal{M}_{3}\cong\textrm{AdS}_3$ and $\cal I$ is spacelike.
A set of global coordinates $\mathrm{(\tau,\rho,\psi,\phi)}$ can be introduced by setting
\begin{align}
   x^{i}=\frac{R}{\cos{\psi}}\,\alpha^{i}\ ,\quad x^{5}=\frac{R}{\cos{\psi}}\,\sin{\psi} 
   \qquad\mathrm{with}\quad \psi\in{\cal I}_\psi\equiv\left(-\tfrac{\pi}{2}, \tfrac{\pi}{2}\right)
   \quad\mathrm{and}\quad \eta_{ij}\alpha^{i}\alpha^{j}=-1\ ,
\end{align}
where $\alpha^{i}=\alpha^{i}(\tau,\rho,\phi)$ for $i=1,\ldots,4$ embeds unit $\mathrm{AdS_{3}}$ into $\mathbb{R}^{2,2}$ with metric $(\eta_{ij})=\mathrm{diag}(-1,-1,1,1)$. A standard parametrization is
\begin{align}
    \alpha^{1}=\cosh{\rho}\,\cos{\tau}\ ,\quad \alpha^{2}=\cosh{\rho}\,\sin{\tau}\ ,\quad
    \alpha^{3}=\sinh{\rho}\,\cos{\phi}\ ,\quad \alpha^{4}=\sinh{\rho}\,\sin{\phi}\ ,
\end{align}
with $\rho\in\R_+$, $\phi\in S^1$ and the temporal coordinate $\tau\in S^1$.
The flat metric on $\mathbb{R}^{2,3}$ then induces a metric on $\mathrm{AdS_{4}}$\;,
\begin{align}\label{metric1}
    \mathrm{d}s^{2}\=\frac{R^{2}}{\cos^{2}\!{\psi}}
    \bigl(\diff\psi^2-\cosh^2\!\rho\,\diff\tau^2+\diff\rho^2+\sinh^2\!\rho\,\diff\phi^2\bigr) \=
    \frac{R^{2}}{\cos^{2}\!{\psi}}\bigl(\mathrm{d}\psi^{2}+\mathrm{d}\Omega^{2}_{1,2}\bigr)\=
    \frac{R^{2}}{\cos^{2}\!{\psi}}\,\mathrm{d}s^{2}_{\mathrm{cyl}}\ ,
\end{align}
where $\mathrm{d}\Omega^{2}_{1,2}$ and $\mathrm{d}s^{2}_{\mathrm{cyl}}$ denote the metric on unit $\mathrm{AdS_{3}}$ and on the finite cylinder $\mathcal{I}_\psi\times \mathrm{AdS_{3}}$, respectively. 

We will take advantage of the fact that
\begin{equation}\label{groupMfd}
\mathcal{M}_{3}\,\cong\,\textrm{AdS}_3\,\cong\,\mathrm{PSL(2,\mathbb{R})} 
\,\cong\,\mathrm{SU(1,1)}/{\{\pm\mathrm{id}\}}\,\cong\,\mathrm{SO}(1,2)
\end{equation}
is a group manifold. In particular, this admits an orthonormal basis $\{e^{\alpha}\}$, $\alpha=0,1,2,$ of SU(1,1) left-invariant one-forms on $\mathrm{AdS_{3}}$, satisfying
\begin{align}
     \mathrm{d}e^{\alpha}+f^{\alpha}_{\ \beta\gamma}\;e^{\beta}\wedge e^{\gamma}=0
\end{align}
where $f_{\ \beta\gamma}^{\alpha}$ are the structure constants of the $\mathfrak{sl}(2,\R)$ Lie algebra. Concretely, $f^{1}_{\ 20}{=}f^{2}_{\ 01}{=}1$ and $f^{0}_{\ 12}{=}{-1}$ with the value of the remaining unrelated constants being zero. This basis can be obtained explicitly by expanding the left Maurer--Cartan one-form
\begin{align}\label{MC1forms}
    \Omega_{L}(g)\=g^{-1}\mathrm{d}g\=e^{\alpha}\,I_{\alpha}\ ,
\end{align}
where
\begin{align}
    I_{0} \= \begin{pmatrix}-\mathrm{i}&0\\0&\mathrm{i}\end{pmatrix}\ ,\qquad 
    I_{1} \= \begin{pmatrix}0&1\\1&0\end{pmatrix}\ ,\qquad 
    I_{2} \= \begin{pmatrix}0&-\mathrm{i}\\\mathrm{i}&0\end{pmatrix}\ ,
\end{align}
are the three $\mathfrak{sl}(2,\R)$ generators subject to
\begin{align}
    [I_{\alpha},I_{\beta}] \= 2\,f_{\ \alpha\beta}^{\gamma}\,I_{\gamma} \quad\und\quad \mathrm{tr}(I_{\alpha}\,I_{\beta}) \= 2\,\eta_{\alpha\beta} \quad\with
    (\eta_{\alpha\beta})=\mathrm{diag}(-1,1,1)\ .
\end{align}
The identification map $g$ is defined as
\begin{align}
    g:\;\mathrm{AdS_{3}}\,\rightarrow\,\mathrm{SU(1,1)}\qquad\mathrm{via}\qquad
    (\alpha^1,\alpha^2,\alpha^3,\alpha^4) \,\mapsto\, 
    \begin{pmatrix} \alpha^1{-}\im\alpha^2 & \alpha^3{-}\im\alpha^4 \\
    \alpha^3{+}\im\alpha^4 & \alpha^1{+}\im\alpha^2 \end{pmatrix}\ .
\end{align}
Using this map the left-invariant one-forms $e^{\alpha}$ \eqref{MC1forms} compute to
\begin{equation}
\begin{aligned}
    e^{0}&\=\cosh^{2}\!{\rho}\;\mathrm{d}\tau+\sinh^{2}\!{\rho}\;\mathrm{d}\phi\ ,\\
    e^{1}&\=\cos{(\tau{-}\phi)}\;\mathrm{d}\rho+\sinh{\rho}\;\cosh{\rho}\;\sin{(\tau{-}\phi)}\;\diff{(\tau{+}\phi)}\ ,\\
    e^{2}&\=-\sin{(\tau{-}\phi)}\;\mathrm{d}\rho+\sinh{\rho}\;\cosh{\rho}\;\cos{(\tau{-}\phi)}\;\diff{(\tau{+}\phi)}\ ,
\end{aligned}
\end{equation}
in terms of which the metric on the finite cylinder $\mathcal{I}_\psi{\times}\mathrm{AdS_{3}}$ is given by
\begin{align}
    \mathrm{d}s^{2}_{\mathrm{cyl}}\=\mathrm{d}\psi^{2}+\eta_{\alpha\beta}\,e^{\alpha}e^{\beta}
    \ =:\ (e^\psi)^2 -(e^{0})^{2}+(e^{1})^{2}+(e^{2})^{2}\ .
\end{align}

\noindent
\textbf{$H^3$- or $S^3_+$-slicing coordinates.} 
Another useful foliation takes $\mathcal{M}_{3}\cong H^3$ and a timelike interval~${\cal I}\ni\tau$ (which in fact is a circle). To exhibit the $H^3$ structure, we change coordinates $(\rho,\psi)\to(\lambda,\theta)\in\R_+{\times}[0,\pi]$ via
\begin{equation}
    \tanh\rho \= \sin\theta\,\tanh\lambda \qquad\mathrm{and}\qquad \tan\psi \= -\cos\theta\,\sinh\lambda \ .
\end{equation}
The full new coordinates $(\tau,\lambda,\theta,\phi)$ can also be obtained by parametrizing the hypersurface in $\R^{2,3}$ as
\begin{align}
    x^{1}=R\cos\tau\,\cosh\lambda\,,\
    x^{2}=R\sin\tau\,\cosh\lambda\,,\
    x^{3}=R\,\beta^1\sinh\lambda\,,\
    x^{4}=R\,\beta^2\sinh\lambda\,,\
    x^{5}=R\,\beta^3\sinh\lambda
\end{align}
where $\beta^{a}=\beta^{a}(\theta,\phi)$ for $a=1,2,3$ embeds unit $S^{2}$ into $\R^3$, i.e.\ $\delta_{ab}\beta^a\beta^b=1$. A standard choice with $\phi\in[0,2\pi)$ is
\begin{align}
\beta^{1}\=\sin\theta\,\cos\phi\ ,\qquad\beta^{2}\=\sin\theta\,\sin\phi\ ,\qquad\beta^{3}\=\cos\theta\ . 
\end{align}
In these coordinates the metric induced on $\mathrm{AdS_{4}}$ takes the form
\begin{equation} \label{metric2}
    \mathrm{d}s^{2}\=R^2\bigl(-\cosh^2\!\lambda\,\diff\tau^2+\diff\lambda^2+\sinh^2\!\lambda\,\diff\Omega_2^2\bigr)
    \qquad\textrm{with}\quad 
    \diff\Omega_2^2=\diff\theta^2+\sin^2\!\theta\diff\phi^2\ ,
\end{equation}
making the hyperbolic space $H^3$ explicit as leaves for $\diff\tau=0$.

For yet another interpretation it is revealing to replace the radial coordinate $\lambda$ by an angle $\chi\in[0,\frac{\pi}{2})$ using
\begin{align}
    \sinh\lambda\=\tan\chi \qquad\Longleftrightarrow\qquad \cosh\lambda\=1/\cos\chi\ .
\end{align}
With this choice the induced metric reads
\begin{align} \label{metric3}
    \mathrm{d}s^{2}\=\frac{R^{2}}{\cos^{2}\!{\chi}}\bigl(-\mathrm{d}{\tau}^{2}+\mathrm{d}\chi^{2}+\sin^{2}\!\chi\,\mathrm{d}\Omega^{2}_{2}\bigr)\=\frac{R^{2}}{\cos^{2}\!{\chi}}\bigl(-\mathrm{d}{\tau}^{2}+\mathrm{d}\Omega^{2}_{3+}\bigr)
\end{align}
where $\mathrm{d}\Omega^{2}_{3+}$ is the round metric on the upper hemisphere $S^{3}_{+}$ of the unit three-sphere $S^{3}=S^{3}_{+}\cup S^{2}\cup S^{3}_{-} $ with the boundary $S^2$ at $\chi{=}\frac{\pi}{2}$. 

Hence, we may view the $H^3$-slice as a hemisphere~$S^3_+$, which shows $\mathrm{AdS_{4}}$ to be conformally equivalent to $S^1\times S^{3}_{+}$. This connects with the construction of SU(2) Yang--Mills solutions on dS$_4\cong\mathcal{I}_\tau\times S^3$ with $\mathcal{I}_\tau=(-\frac{\pi}{2},\frac{\pi}{2})$ performed in~\cite{IvLePo1,IvLePo2}. However, unless we pass to the universal covering $\widetilde{\textrm{AdS}}_4$, the periodicity in~$\tau$ severely restricts the existence of such solutions on AdS$_4$. Nevertheless, we can recycle the map of~\cite{zhilin,kumar} from $\mathcal{I}_\tau\times S^3$ to Minkowski space by restricting it to a hemisphere.

%~~~~~~~~~~~~~~~~~~~~~~~~~~~~~~~~~~~~~~~~~~~~~~~~~~~
\newpage

\section{SU(1, 1) Yang--Mills fields on $\text{AdS}_{4}$}
\noindent
Since AdS$_4$ is conformally flat, a solution of the Yang--Mills equations on this spacetime will also yield
\begin{wrapfigure}{r}{0.25\textwidth}
\centering
\vspace{-3mm}
\includegraphics[width=0.25\textwidth]{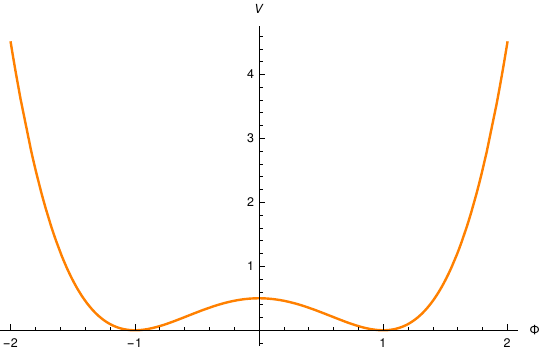}
\caption{Potential $V(\Phi)$.}
\label{potential}
\end{wrapfigure}
Yang--Mills fields on Minkowski space. In the AdS$_3$ coordinates~$(\tau,\rho,\psi,\phi)$ such solutions are easy to come by for an SU(1,1) gauge group, because the metric~\eqref{metric1} shows that a $\psi$-foliation has the $\mathrm{SU}(1,1)/\{\pm\mathrm{id}\}$ group manifold~\eqref{groupMfd} as leaves.
We therefore consider a gauge connection $\mathcal{A}$ and its curvature $\mathcal{F}=\mathrm{d}\mathcal{A}+\mathcal{A}\wedge\mathcal{A}$ taking values in the $\mathfrak{su}(1,1)$ Lie algebra, which in the orthonormal frame $\{e^{\psi},e^{\alpha}\}$ can be expressed as
\begin{align}
    \mathcal{A}\= \mathcal{A}_{\psi}\,e^{\psi}+ \mathcal{A}_{\alpha}\,e^{\alpha}
    \quad\implies\quad
    \mathcal{F}\=\mathcal{F}_{\psi\alpha}\;e^{\psi}\wedge e^{\alpha}+\tfrac{1}{2}\mathcal{F}_{\beta\gamma}\;e^{\beta}\wedge e^{\gamma}\ .
\end{align}
The SU(1,1) symmetry suggests a gauge fixing $\mathcal{A}_{\psi}=0$ and the ansatz~\cite{IvLePo2}
\begin{align}
    \mathcal{A}\=X_{\alpha}(\psi)\;e^{\alpha}
    \qquad\implies\qquad
    \mathcal{F}\=X_{\alpha}^{'}\;e^{\psi}\wedge e^{\alpha}+\tfrac{1}{2}\bigl(-2f_{\ \beta\gamma}^{\alpha}X_{\alpha}+[X_{\beta},X_{\gamma}]\bigr)\;e^{\beta}\wedge e^{\gamma}
\end{align}
where $X_{\alpha}^{'}{:=}\frac{\diff X\alpha}{\diff\psi}$ and the three functions $X_{\alpha}$ of $\psi\in{\cal I}$ alone take values in the $\mathfrak{su}(1,1)$ Lie algebra. In terms of the matrix functions $X_{\alpha}$ the Yang--Mills Lagrangian on $\mathcal{I}_\psi\times\mathrm{AdS_{3}}$ is then given by
\begin{equation} \label{matrixlag}
\begin{aligned}
    \mathcal{L}&\=\tfrac14\mathrm{tr}\mathcal{F}_{\psi\alpha}\mathcal{F}^{\psi\alpha}+\tfrac18\mathrm{tr}\mathcal{F}_{\beta\gamma}\mathcal{F}^{\beta\gamma}\\[4pt]
    &\=\mathrm{tr}\Bigl\{\tfrac14\bigl((X_{1}^{'})^{2}+(X_{2}^{'})^{2}-(X_{0}^{'})^{2}\bigr)-\bigl(X_{1}-\tfrac12[X_{2},X_{0}]\bigr)^{2}-\bigl(X_{2}-\tfrac12[X_{0},X_{1}]\bigr)^{2}+\bigl(X_{0}+\tfrac12[X_{1},X_{2}]\bigr)^{2}\Bigr\}\ .
\end{aligned} 
\end{equation}
In order to satisfy the additional Gauss-law constraint $[X_{\alpha},X^{'}_{\alpha}]=0$, we further specialize $X_\alpha=X_\alpha^\beta I_\beta$ to
\begin{align} \label{Thetaansatz}
    X_\alpha^\beta = \Theta_\alpha(\psi)\,\delta_\alpha^\beta
    \qquad\Longleftrightarrow\qquad
    X_{0}=\Theta_{0}\,{I_{0}}\ ,\quad 
    X_{1}=\Theta_{1}\,{I_{1}}\ ,\quad
    X_{2}=\Theta_{2}\,{I_{2}}\ ,
\end{align}
where $\Theta_{\alpha}$ are real functions of $\psi\in\mathcal{I}$.
Gauge potential and field strength then take the form
\begin{equation}
    \mathcal{A}\= \Theta_\alpha\,I_\alpha\,e^\alpha
    \qquad\Longrightarrow\qquad
    \mathcal{F}\= \Theta'_\alpha\,I_\alpha\,e^\psi\wedge e^\alpha\ -\ (\Theta_\alpha{-}\Theta_\beta\Theta_\gamma)\,I_\alpha\,f^\alpha_{\ \beta\gamma}\,e^\beta\wedge e^\gamma\ .
\end{equation}
Our ansatz allows us to perform the trace in~\eqref{matrixlag} and obtain
\begin{equation} \label{Thetalag}
    \mathcal{L} \= \tfrac12\bigl\{(\Theta'_1)^2+(\Theta'_2)^2+(\Theta'_0)^2\bigr\}\ -\ 
    2\bigl\{(\Theta_1{-}\Theta_2\Theta_0)^2+(\Theta_2{-}\Theta_0\Theta_1)^2+(\Theta_0{-}\Theta_1\Theta_2)^2\bigr\}\ .
\end{equation}
We note that, contrary to how the SU(2) Lagrangian on AdS$_4$ worked out in~\cite{IvLePo2}, this one is identical to the SU(2) Lagrangian on dS$_4$ obtained there.  
We further distinguish between two cases:

\noindent
\textbf{Non-Abelian SU(1,1) equivariant ansatz.} 
Equivariance requires $\mathcal{A}\propto I_\alpha\,e^\alpha$~\cite{uenal} and thus
\begin{align} \label{3.9}
    \Theta_{0}=\Theta_{1}=\Theta_{2}\ =:\ \tfrac{1}{2}\bigl(1+\mathrm{\Phi(\psi)}\bigr)\ ,
\end{align}
where we have parametrized the single free function of $\psi\in\mathcal{I}$ in a convenient way. Because all three $\mathfrak{su}(1,1)$ generators are excited, the field configuration is non-Abelian and takes the form
\begin{align} \label{3.15}
    \mathcal{A}\=\tfrac{1}{2}\bigl(1+\Phi(\psi)\bigr)\,I_{\alpha}\,e^{\alpha}
    \qquad\implies\qquad
    \mathcal{F}\=\tfrac{1}{2}\,\Phi'(\psi)\,I_{\alpha}\,e^{\psi}\wedge e^{\alpha}\ -\ \tfrac{1}{4}\bigl(1{-}\Phi(\psi)^{2}\bigr)\,I_{\alpha}\,f^{\alpha}_{\ \beta\gamma}\,e^{\beta}\wedge e^{\gamma}
\end{align}
giving rise to a Lagrangian~(cf.~\cite{IvLePo1})
\begin{equation}
    \mathcal{L}\=\tfrac34\,\Bigl\{ \tfrac12(\Phi^{'})^{2}\ -\ \tfrac12(1-\Phi^{2})^{2} \Bigr\}
\end{equation}
for the single degree of freedom~$\Phi$, whose equation of motion reads
\begin{align} \label{Phieom}
    \Phi^{''}\=2\,\Phi\,(1-\Phi^{2})\= -\tfrac{\pa V}{\pa \Phi} 
    \qquad\quad\textrm{where}\qquad V(\Phi)\=\tfrac{1}{2}(\Phi^2-1)^2\ . 
\end{align}
Apparently, the degree of freedom~$\Phi$ behaves just like the coordinate of a Newtonian particle moving in  
the interval $\left(-\frac{\pi}{2},\frac{\pi}{2}\right)$ under the influence of a double-well potential $V(\Phi)$ of Figure \ref{potential}. The solutions to the equation of motion~\eqref{Phieom} are well-known in terms of Jacobi elliptic functions.

From \eqref{3.15} we read off the color-electric and color-magnetic fields as \\ $\mathcal{E_{\alpha}}\equiv\mathcal{F_{\alpha\psi}}=\tfrac{1}{2}{\Phi'(\psi)}\;I_{\alpha}$ \ and \ $\mathcal{B_{\alpha}}\equiv\tfrac{1}{2}f_{\alpha}^{\ \beta\gamma}\mathcal{F_{\beta\gamma}}=\tfrac{1}{2}\bigl(1-\Phi(\psi)^{2}\bigr)\,I_{\alpha}$ \
respectively in the \enquote{$\mathrm{AdS}_{3}$-frame}, where indices on the structure constants have been raised and lowered using the metric~$(\eta_{\alpha\beta})$.
The action of these field configurations is infinite, since the integral over the AdS$_3$ slice produces its infinite volume.

The energy-momentum tensor of these Yang--Mills fields in the orthonormal frame can be computed straightforwardly via the expressions
\begin{equation}
    \mathcal{T}_{\alpha\beta}=-\tfrac{1}{2g^2}\mathrm{tr}\bigl\{\mathcal{F}_{\alpha\psi}\mathcal{F}_{\beta\psi}+\mathcal{F}_{\alpha\gamma}\mathcal{F}_{\beta\delta}\,\eta^{\gamma\delta}-\tfrac14\eta_{\alpha\beta}\mathcal{F}^2\bigr\}
    \qquad\mathrm{and}\qquad
    \mathcal{T}_{\psi\psi}=-\tfrac{1}{2g^2}\mathrm{tr}\bigl\{\mathcal{F}_{\psi\gamma}\mathcal{F}_{\psi\delta}\,\eta^{\gamma\delta}-\tfrac14\mathcal{F}^2\bigr\} 
\end{equation}
with \ $\mathcal{F}^2\equiv2\mathcal{F}_{\psi\delta}\mathcal{F}^{\psi\delta}+\mathcal{F}_{\gamma\delta}\mathcal{F}^{\gamma\delta}$ \ and \ $\mathcal{T}_{\alpha\psi}=0$.
Explicitly one finds (ordering $\alpha,\beta=0,1,2$)
\begin{align} \label{TAdS}
    \begin{pmatrix}\mathcal{T}_{\alpha\beta} & \mathcal{T}_{\alpha\psi} \\ \mathcal{T}_{\psi\beta} & \mathcal{T}_{\psi\psi} \end{pmatrix} \= 
    \tfrac{1}{2g^{2}}\left(\begin{smallmatrix} \rho & 0 & 0 & 0\\ 0 & -\rho & 0 & 0\\0 & 0 & -\rho & 0\\0 & 0 & 0 & 3\rho \end{smallmatrix}\right) 
    \ \ \with
    \rho=-\tfrac{1}{2}\mathrm{tr}(\mathcal{E}_{\alpha}\mathcal{E}_{\alpha}{+}\mathcal{B}_{\alpha}\mathcal{B}_{\alpha})
    =-\tfrac14\bigl((\Phi')^2+(1{-}\Phi^2)^2\bigr)\ .
\end{align}
It is traceless as expected, but the conserved energy density $\rho:=2g^2\,\mathcal{T}_{00}$ is negative 
since the noncompact generators dominate the sum.

\noindent
\textbf{Abelian (non-equivariant) ansatz.} 
The ansatz~\eqref{Thetaansatz} always includes Abelian configurations by putting to zero two of the three functions~$\Theta_\alpha$. Let us keep the 0-axis in isospace (the compact generator) and set
\begin{align}
    \Theta_{0}=\mathrm{h}(\psi)\qquad\mathrm{while}\qquad \Theta_{1}= \Theta_{2}=0\ ,
\end{align}
where $\mathrm{h}(\psi)$ is some real-valued function of $\psi \in\mathcal{I}_\psi$. Dropping the single generator by writing $\mathcal{A}=\widetilde{\mathcal{A}}\,I_0$ the Abelian configuration takes the form~\footnote{We need to be careful in the following with a minus sign arising from $(I_0)^2$.}
\begin{equation}
    \widetilde{\mathcal{A}}\= \mathrm{h}(\psi)\,e^0 \qquad\Longrightarrow\qquad
    \widetilde{\mathcal{F}}\= \mathrm{h}'(\psi)\,e^\psi\wedge e^0\ +\ 2\,\mathrm{h}(\psi)\,e^1\wedge e^2\ ,
\end{equation}
hence only $\widetilde{\mathcal{E}}_0$ and $\widetilde{\mathcal{B}}_0$ appear.
The Lagrangian~\eqref{Thetalag} reduces to
\begin{align}
    \mathcal{L} \= \tfrac12\,(\mathrm{h}')^2\ -\ 2\,\mathrm{h}^2
    \qquad\Longrightarrow\qquad
    \mathrm{h}''=-4\,\mathrm{h}\ .
\end{align}
The same happens for the other choices, mimicking a particle in a harmonic potential. The general solution may be expressed as
\begin{align} \label{anharEOM}
    \mathrm{h}(\psi)\=-\tfrac12\,f\,\cos2(\psi{-}\psi_{0})
\end{align}
with two parameters $f$ and $\psi_{0}$. The corresponding gauge potential and field strength read
\begin{align} \label{abelianAF}
    \mathcal{\widetilde{A}}\=-\tfrac12\,f\,\cos2(\psi{-}\psi_{0})\,e^{0}
    \qquad\mathrm{and}\qquad
    \mathcal{\widetilde{F}}\= f\,\bigl\{ \sin2(\psi{-}\psi_{0})\,e^{\psi}\wedge e^{0}-\cos2(\psi{-}\psi_{0})\,e^{1}\wedge e^{2}\big\}\ .
\end{align}
Here, the action integral is indeterminate: 
integrating over the AdS$_3$ slice still yields its infinite volume, 
but kinetic and potential contributions cancel in the $\psi$ integration
since this system is a harmonic oscillator.
Finally, using $\tr(I_0^2)=-2$ the traceless energy-momentum tensor is given by
\begin{align}
    \begin{pmatrix}\widetilde{\mathcal{T}}_{\alpha\beta} & \widetilde{\mathcal{T}}_{\alpha\psi} \\ \widetilde{\mathcal{T}}_{\psi\beta} & \widetilde{\mathcal{T}}_{\psi\psi} \end{pmatrix} \= 
    \tfrac{1}{2g^{2}}\, \left(\begin{smallmatrix} \tilde{\rho} & 0 & 0 & 0\\ 0 & \tilde{\rho} & 0 & 0\\0 & 0 & \tilde{\rho} & 0\\0 & 0 & 0 & -\tilde{\rho} \end{smallmatrix}\right)
    \qquad\mathrm{with}\quad \tilde{\rho}=(\mathrm{h}')^2+4\mathrm{h}^2=f^2\ .
\end{align}
This form represents a well-known non-null electrovacuum configuration.

%~~~~~~~~~~~~~~~~~~~~~~~~~~~~~~~~~~~~~~~~~~~~~~~~~~~
\newpage

\section{Push forward of solutions to Minkowski space}
\noindent
\textbf{Mapping to Minkowski space.} 
In the previous section we solved the vacuum $\mathrm{SU(1,1)}$ Yang--Mills equations on the cylinder $\mathcal{I}_\psi\times \mathrm{AdS_{3}}$. We are now interested in carrying these solutions to the conformally related Minkowski space and analyzing the properties of the solutions on~$\R^{1,3}$. 
To achieve this, we need to construct the conformal map from the AdS$_3$ slicing coordinates 
$(\tau,\rho,\psi,\phi)$ to the Minkowski coordinates 
$(t,x,y,z)=(t,x^a)=(x^\mu)$ with $a=1,2,3$ and $\mu=0,1,2,3$.
We do this in two steps, keeping $\phi$ fixed. 
Firstly, keeping $\tau$ constant we pass to the intermediate $H^3$ (or $S^3_\pm$) slicing coordinates $(\tau,\lambda\,\mathrm{or}\,\chi,\theta,\phi)$ via
\begin{align} \label{firstmap}
   \tanh{\rho} = \sin\theta\,\tanh\lambda = \varepsilon\,\sin{\theta}\,\sin{\chi}
   \qquad\mathrm{and}\qquad
   \tan{\psi} = -\varepsilon\,\cos\theta\,\sinh\lambda = -\varepsilon\,\cos{\theta}\,\tan{\chi}\ .
\end{align}
The two choices $\varepsilon=\pm1$ here correspond to two different versions of the map:
\begin{equation}
\begin{aligned}
    \varepsilon=+1\ &:\quad \rho,\lambda\in\R_+\,,\ \chi\in[0,\tfrac{\pi}{2}) 
    \quad\Leftrightarrow\quad \textrm{northern hemisphere}\ S^3_+\\
    \varepsilon=-1\ &:\quad \rho,\lambda\in\R_-\,,\ \chi\in(\tfrac{\pi}{2},\pi]
    \quad\Leftrightarrow\quad \textrm{southern hemisphere}\ S^3_-
\end{aligned}
\end{equation}
where $\psi\in(-\tfrac{\pi}{2},\tfrac{\pi}{2})$ and $\theta\in[0,\pi]$ throughout.
Both versions will be needed since either one covers only half of Minkowski space as we shall see.

Secondly, we map to Minkowski coordinates $(t,x,y,z)$ or $(t,r,\theta,\phi)$ related via
\begin{equation}
    (x,y,z)\=(r\,\sin\theta\,\cos\phi,\ r\,\sin\theta\, \sin\phi,\ r\,\cos\theta) 
    \qquad\Longrightarrow\qquad r^{2}=x^{2}+y^{2}+z^{2}
\end{equation}
by the following relations~\cite{zhilin,kumar}:
\begin{align} \label{Mmap1}
    \cos\tau\=\frac{\gamma}{2R^2}\,\bigl(r^{2}-t^{2}+R^{2}\bigr)
    \qquad\mathrm{and}\qquad 
    \cos\chi\=\frac{\gamma}{2R^2}\,\bigl(r^{2}-t^{2}-R^{2}\bigr)
\end{align}
or
\begin{align} \label{Mmap2}
    \sin\tau\=\gamma\,t/R
    \qquad\mathrm{and}\qquad
    \sin\chi\=\gamma\,r/R
    \qquad\implies\qquad
    \frac{\sin\chi}{\sin\tau}\=\frac{r}{t}
\end{align}
where 
\begin{align} \label{Mgamma}
    \gamma\=\frac{2R^{2}}{\sqrt{4R^{2}t^{2}+(r^{2}-t^{2}+R^{2})^{2}}}
    \=\frac{2R^{2}}{\sqrt{4R^{2}r^{2}+(r^{2}-t^{2}-R^{2})^{2}}}\ .
\end{align}
The $S^2$ coordinates $(\theta,\phi)$ are identified on both sides of this map. It follows that
\begin{align} \label{chitau}
    \gamma\=\cos{\tau}-\cos{\chi}\ >0
    \qquad\implies\qquad\chi>\lvert\tau\rvert
\end{align}
and therefore the inverse relations
\begin{align}
    \frac{t}{R}\=\frac{\sin{\tau}}{\cos{\tau}-\cos{\chi}}
    \qquad\mathrm{and}\qquad
    \frac{r}{R}\=\frac{\sin{\chi}}{\cos{\tau}-\cos{\chi}}\ .
\end{align}
The inequality in~\eqref{chitau} tells us that only half of the AdS domain is mapped into Minkowski space,
because the lines $\chi{=}\lvert\tau\rvert$ correspond to null infinity.
We can take this into account by restricting
\begin{equation}
    \cos\tau\ >\ \mathrm{sgn}(\rho)\sqrt{\tfrac{1-\tanh^2\!\rho}{1+\tan^2\!\psi}}\ .
\end{equation}

\begin{figure}[ht!]
    \centering
    \includegraphics[height=6cm,width=4.5cm]{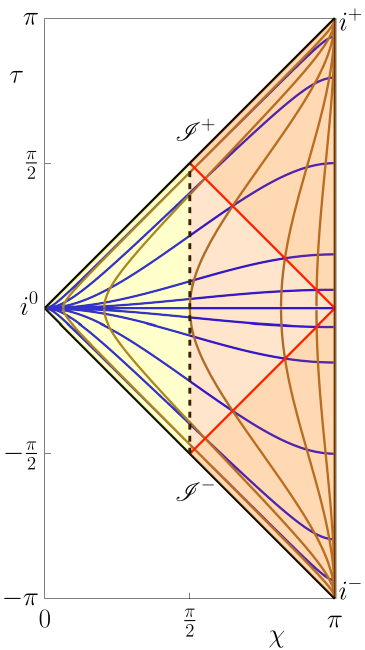}
    \qquad\qquad\qquad
    \includegraphics[height=6cm,width=4cm]{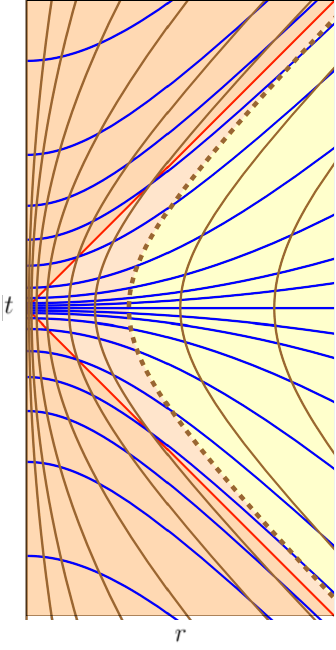}
    \caption{Gluing of two $\textrm{AdS}_4$ to reveal the full Minkowski space with the lightcone (red). Left: $(\tau,\chi)$ AdS$_4$ space (two copies) yielding the Penrose diagram with constant $t$- (blue) and $r$-slices (brown). Right: $(t,r)$ Minkowski space with boundary hyperbola $H^{1,2}_R$ (dashed) and constant $\tau$- (blue) and $\chi$-slices (brown).}
    \label{MinkDiag}
\end{figure}

A direct relation between AdS$_3$-slicing and Minkowski coordinates reads
\begin{equation} \label{AdStoMink}
    \tanh\rho \= 2\varepsilon\,R\,\sqrt{\frac{\phantom{I}\smash{r^2-z^2}}{4R^2r^2+(r^2{-}t^2{-}R^2)^2}}
    \qquad\und\qquad
    \tan\psi \= \frac{-2\varepsilon\,R\,z}{r^2{-}t^2{-}R^2} \ ,
\end{equation}
and the AdS$_4$ metric \eqref{metric1}, \eqref{metric2} or~\eqref{metric3} becomes
\begin{equation} \label{metric4}
    \diff s^2 \= \frac{4\,R^4}{(r^2{-}t^2{-}R^2)^2}\,\bigl(-\diff t^2+\diff x^2+\diff y^2+\diff z^2\bigr)
    \= \frac{4\,R^4}{(r^2{-}t^2{-}R^2)^2}\,\bigl(-\diff t^2+\diff r^2+r^2\diff\Omega_2^2\bigr)\ ,
\end{equation}
where the conformal factor diverges at the boundary
\begin{equation}
    \bigl\{|\psi|{=}\tfrac{\pi}{2}\bigr\} \= \bigl\{\lambda{=}{\pm}\infty\bigr\}
    \= \bigl\{\chi{=}\tfrac{\pi}{2}\bigr\} \qquad\Longleftrightarrow\qquad
    \bigl\{r^2{-}t^2{=}R^2\bigr\} \ =:\ H^{1,2}_R\ \cong\ \mathrm{dS}_3\ 
    \cong\ \mathcal{I}_\tau{\times}S^2\big|_\textrm{bdy}\ .
\end{equation}
The ``northern map'' ($\varepsilon{=}{+}1$ and $\chi\in[0,\tfrac{\pi}{2})$) covers the
exterior of the one-sheeted hyperboloid~$H^{1,2}_R$, while the ``southern map'' ($\varepsilon{=}{-}1$ and $\chi\in(\tfrac{\pi}{2},\pi]$) yields the interior.
Hence, all of Minkowski space is covered for $\tau\in(-\pi,\pi)$ compatible with one time-circle of~AdS$_4$
and for $\chi\in[0,\pi]$.
Indeed, the north pole ($\chi{=}0$) is mapped to spatial infinity $(t,r{=}\infty)$ while the south pole ($\chi{=}\pi$) lands at the spatial origin $(t,r{=}0)$.\footnote{
Note that $\lambda{=}0$ corresponds to the north pole in the $\varepsilon{=}{+}1$ version but to the south pole in the $\varepsilon{=}{-}1$ version of the map. More generally, the $\rho{=}0$ line maps to different $\chi$ segments at $\theta{=}0$ or $\pi$ in the two versions.} This is clearly demonstrated in Figure \ref{MinkDiag} where the domain $(\tau,\chi)$ is revealed as the Penrose diagram of Minkowski space.

\begin{figure}[h!]
    \centering
    \includegraphics[height=3cm,width=15cm,trim={0.5cm 19.5cm 0.5cm 4cm},clip]{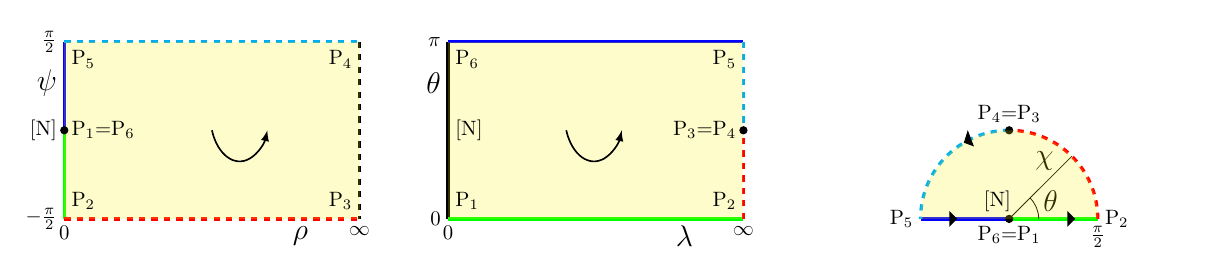}
    \caption{Illustration of the boundary of $H^3_+\cong S^3_+$ in three different coordinate systems
    with color-coded segments generated by points $\mathrm{P}_1,\ldots,\text{P}_6$ and containing the north pole [N].}
    \label{bdyLeft}
\end{figure}

\begin{figure}[h!]
    \centering
    \includegraphics[height=3cm,width=15cm,trim={0.5cm 13.5cm 0.5cm 10cm},clip]{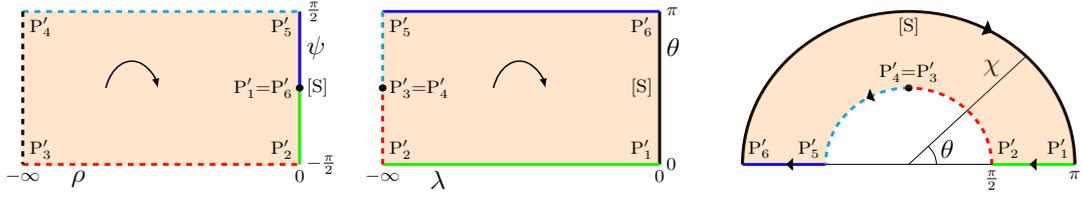}
    \caption{Illustration of the boundary of $H^3_-\cong S^3_-$ in three different coordinate systems
    with color-coded segments generated by points $\text{P}'_1,\ldots,\text{P}'_6$ and containing the south pole [S].}
    \label{bdyRight}
\end{figure}

\begin{figure}[h!]
    \centering
    \includegraphics[height=3cm,width=15cm,trim={0.5cm 7.8cm 0.5cm 15.8cm},clip]{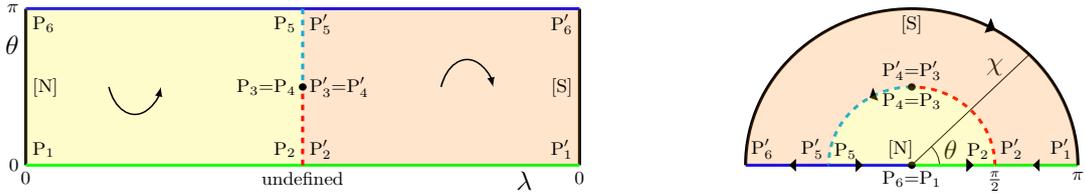}
    \caption{Gluing $S_3^+$ (yellow shaded region) to $S_3^-$ (orange shaded region) along the (dashed) boundary.}
    \label{bdyGlued}
\end{figure}

Some attention has to be payed to the gluing (for fixed~$\tau$) of the two copies $H^3_\pm\cong S^3_\pm$ along their boundaries $\partial S^3_\pm=S^2\big|_\textrm{bdy}$
parametrized by $\{\theta,\phi\}$.
We have chosen the signs in \eqref{firstmap} such that the boundary orientations are not altered by the maps and are compatible with the gluing. The two versions of the maps $(\psi,\rho)\rightarrow(\lambda,\theta)\rightarrow(\chi,\theta)$ are evaluated for some key points in Tables \ref{table1} and \ref{table2} (with $\epsilon{>}0$ infinitesimal) and illustrated in Figures \ref{bdyLeft} and \ref{bdyRight} below
where dashed lines represent the $S^3_\pm$ boundary while solid lines lie in the interior, running from a pole to the boundary and back.
Equally colored lines are identified by the maps (and on the boundary also by gluing). Note that the maps identify $\mathrm{P}_1{\equiv}\mathrm{P}_6$ (the north pole) and $\mathrm{P}_3{\equiv}\mathrm{P}_4$ (on the boundary), and likewise for the southern copy. The two hemispheres are glued by identifying the points of the pairs $(\mathrm{P_{2}},\mathrm{P}'_2)$, $(\mathrm{P_{3}},\mathrm{P}'_3)$, $(\mathrm{P_{4}},\mathrm{P}'_4)$,  and $(\mathrm{P_{5}},\mathrm{P}'_5)$ 
to obtain the entire $S^{3}=S^{3}_{+}\cup S^{2}\cup S^{3}_{-}$ as depicted in Figure \ref{bdyGlued}.
\begin{table}[h!]
\parbox{.45\linewidth}{
\centering
\begin{tabular}{|c|c|c|c|}
    \hline   & $(\rho,\psi)$ & $(\lambda,\theta)$ & $(\chi,\theta)$ \\ \hline 
    $\mathrm{P}_{1}$ & $(0,0{-}\epsilon)$ & $(0,0)$ & $(0,0)$ \\ 
    $\mathrm{P}_{2}$ & $(0,-\frac{\pi}{2})$ & $(\infty,0)$ & $(\frac{\pi}{2},0)$ \\  
    $\mathrm{P}_{3}$ & $(\infty,-\frac{\pi}{2})$ & $(\infty,\frac{\pi}{2}{-}\epsilon)$ & $(\frac{\pi}{2},\frac{\pi}{2}{-}\epsilon)$\\
    $\mathrm{P}_{4}$ & $(\infty,\frac{\pi}{2})$ & $(\infty,\frac{\pi}{2}{+}\epsilon)$ & $(\frac{\pi}{2},\frac{\pi}{2}{+}\epsilon)$ \\  
    $\mathrm{P}_{5}$ & $(0,\frac{\pi}{2})$ & $(\infty,\pi)$ & $(\frac{\pi}{2},\pi)$\\   
    $\mathrm{P}_{6}$ & $(0,0{+}\epsilon)$ & $(0,\pi)$ & $(0,\pi)$\\ \hline
\end{tabular}
\caption{Key points on the northern copy $H^3_+\cong S^3_+$ in three coordinate systems.\label{table1}}
}
\hfill
\parbox{.45\linewidth}{
\centering
\begin{tabular}{|c|c|c|c|}
    \hline   & $(\rho,\psi)$ & $(\lambda,\theta)$ & $(\chi,\theta)$ \\ \hline 
    {$\mathrm{P}'_1$} & $(0,0{-}\epsilon)$ & $(0,0)$ & $(\pi,0)$ \\  
    {$\mathrm{P}'_2$} & $(0,-\frac{\pi}{2})$ & $(-\infty,0)$ & $(\frac{\pi}{2},0)$ \\   
    {$\mathrm{P}'_3$} & $(-\infty,-\frac{\pi}{2})$ & $(-\infty,\frac{\pi}{2}{-}\epsilon)$ & $(\frac{\pi}{2},\frac{\pi}{2}{-}\epsilon)$ \\  
    {$\mathrm{P}'_4$} & $(-\infty,\frac{\pi}{2})$ & $(-\infty,\frac{\pi}{2}{+}\epsilon)$ & $(\frac{\pi}{2},\frac{\pi}{2}{+}\epsilon)$ \\   
    {$\mathrm{P}'_5$} & $(0,\frac{\pi}{2})$ & $(-\infty,\pi)$ & $(\frac{\pi}{2},\pi)$ \\
    {$\mathrm{P}'_6$} & $(0,0{+}\epsilon)$ & $(0,\pi)$ & $(\pi,\pi)$ \\ \hline
\end{tabular}
\caption{Key points on the southern copy $H^3_-\cong S^3_-$ in three coordinate systems.\label{table2}}
}
\end{table}
\clearpage

\noindent
\textbf{Mapping the one-forms.} 
We first rewrite the SU(1,1) left-invariant one-forms $\{e^{\alpha},e^{\psi}\}$ in terms of the intermediate coordinates $(\tau,\chi,\theta,\phi)=:(y^m)$ for $m=0,1,2,3$.
A direct calculation using \eqref{firstmap} yields
\begin{equation}\label{4.2}
\begin{aligned}
e^{0}&\=\frac{1}{1-\sin^{2}\!\chi\sin^{2}\!{\theta}}\;
\big(\mathrm{d}\tau+\sin^{2}\!{\chi}\,\sin^{2}\!{\theta}\,\mathrm{d}\phi\big)\ ,\\
e^{1}&\=\frac{\varepsilon}{1-\sin^{2}\!{\chi}\sin^{2}\!{\theta}}\;
\big(\cos(\tau{-}\phi)\,(\cos\chi\,\sin\theta\,\mathrm{d}\chi+\sin\chi\,\cos\theta\,\mathrm{d}\theta)+\sin\chi\,\sin\theta\,\sin(\tau{-}\phi)\,\mathrm{d}(\tau{+}\phi)\big)\ ,\\
e^{2}&\=\frac{-\varepsilon}{1-\sin^{2}\!{\chi}\sin^{2}\!{\theta}}\;\big(\sin(\tau{-}\phi)\,(\cos\chi\,\sin\theta\,\mathrm{d}\chi+\sin\chi\,\cos\theta\,\mathrm{d}\theta)-\sin\chi\,\sin\theta\,\cos(\tau{-}\phi)\,\mathrm{d}(\tau{+}\phi)\big)\ ,\\
e^{\psi}=:e^{3}&\=\frac{-\varepsilon}{1-\sin^{2}\!{\chi}\sin^{2}\!{\theta}}\;\big(\cos\theta\,\mathrm{d}\chi-\sin\chi\,\cos\chi\,\sin\theta\,\mathrm{d}\theta\big)\ .
\end{aligned}
\end{equation}
In other words, we know $e^\alpha=e^\alpha_{\ m}\diff y^m$ and $e^\psi=e^\psi_{\ m}\diff y^m$.
Once we include $\chi{=}\frac{\pi}{2}$, the one-forms develop a coordinate singularity at the equator of the boundary two-sphere, $\chi{=}\theta{=}\frac{\pi}{2}$. As we shall see later, this singularity shows up in the Minkowski coordinates as well, at the intersection of $H^{1,2}_R$ with $\R^{1,2}\cong\{z{=}0\}$. 

Our next step is to write down these one-forms in terms of the Minkowski coordinates, which can easily be done via the above map \eqref{Mmap1}--\eqref{Mgamma}.
Without loss of generality we set the scale factor $R{=}1$ for the ease of calculations. It can be recovered anytime  by simply setting $\{x^{\mu}\}\rightarrow\{\frac{x^{\mu}}{R}\}$. 
The Jacobian of the transformations is then obtained as~\cite{kumar}
\begin{equation} \label{Jacobian}
\bigl(J^m_{\ \ \mu}\bigr) \ :=\ \frac{\pa(\tau,\chi,\theta,\phi)}{\pa(t,r,\theta,\phi)}
\= \biggl(\begin{matrix} p & -q \\[4pt] q & -p \end{matrix} \biggr) \oplus\mathds{1}_2
\quad\with\biggl\{ \begin{array}{l}
p\=\gamma^2 (r^2{+}t^2{+}1)/2\=1{-}\cos\tau\cos\chi \\[4pt]
q\=\gamma^2\,t\,r\=\sin\tau\sin\chi \end{array}
\end{equation}
or, choosing Cartesian Minkowski coordinates,
\begin{align}
    \bigl(J^m_{\ \ \mu}\bigr) \ :=\frac{\partial(\tau,\chi,\theta,\phi)}{\partial(t,x,y,z)} \= 
    \begin{pmatrix}
    p & -\frac{x}{r}\,q & -\frac{y}{r}\,q & -\frac{z}{r}\,q \\[4pt]
    q & -\frac{x}{r}\,p & -\frac{y}{r}p & -\frac{z}{r}p \\[2pt]
    0 & \frac{z\,x}{r^{2}\sqrt{x^{2}+y^{2}}} & \frac{z\,y}{r^{2}\sqrt{x^{2}+y^{2}}} & -\frac{\sqrt{x^{2}+y^{2}}}{r^{2}} \\[2pt]
    0 & -\frac{y}{x^{2}+y^{2}} & \frac{x}{x^{2}+y^{2}} & 0  
    \end{pmatrix}\ .
\end{align}
A straightforward but lengthy calculation then yields 
\begin{equation}
    e^\alpha\=e^\alpha_{\ m}J^m_{\ \ \mu}\,\diff x^\mu\ =:\ e^\alpha_{\ \mu}\,\diff x^\mu \quad\und\quad
    e^\psi\=e^\psi_{\ m}J^m_{\ \ \mu}\,\diff x^\mu\ =:\ e^\psi_{\ \mu}\,\diff x^\mu\ ,
\end{equation}
which, using the Minkowski product $x\cdot \diff{x} := x_\mu\diff{x}^\mu$ and $\varepsilon^0:=1$, $\varepsilon^1=\varepsilon^2:=\varepsilon$, becomes
\begin{equation}\label{1formsMink}
\begin{aligned}
    e^{\alpha}&\= \sfrac{\varepsilon^\alpha}{\lambda^2{+}4z^2}\,\big(2(\lambda{+}2)\,\diff{x}^\alpha - 4x^\alpha\,x\cdot\diff{x} - 4\,f^\alpha_{\ \beta\gamma}\,x^\beta\diff{x}^\gamma \big)\ , \\
    e^{\psi}&\=\sfrac{\varepsilon}{\lambda^2{+}4z^2}\,\big({-}2\lambda\,\diff{z} + 4z\,x\cdot\diff{x} \big)\ ,\qquad \textrm{where}\quad \lambda\ :=\ r^2-t^2-1\ .
\end{aligned}
\end{equation}
%~~~~~~~~~~~~~~~~~~~~~~~~~~~~~~~~~~~~~~~~~~~~~~~~~~~

\section{Exact Minkowskian gauge fields}

\noindent
\textbf{Non-Abelian solutions.}
It is now a straightforward exercise to pull the $\mathrm{SU(1,1)}$-equivariant gauge potential $\mathcal{A}$ and its field strength $\mathcal{F}$ \eqref{3.15} over to Minkowski space via
\begin{equation}
\begin{aligned}
    \mathcal{A}\equiv A&\=\tfrac{1}{2}\Big(1+\Phi\bigl(\psi(x)\bigr)\Big)\,I_{\alpha}\;e^{\alpha}_{\ \mu}\;\mathrm{d}x^{\mu} \quad\und \\
    \Fcal \equiv F&\=
    \tfrac{1}{2}\Big(\Phi'\bigl(\psi(x)\bigr)\,I_{\alpha}\,e^{\psi}_{\ \mu}e^{\alpha}_{\ \nu}\ -\ \tfrac{1}{2}\bigl(1{-}\Phi\bigl(\psi(x)\bigr)^{2}\bigr)\,I_{\alpha}\,f^{\alpha}_{\ \beta\gamma}\,e^{\beta}_{\ \mu}e^{\gamma}_{\ \nu} \Big)\,\diff x^\mu\wedge\diff x^\nu\ .
\end{aligned}
\end{equation}
Notice that the gauge fixing $\mathcal{A}_\psi{=}0$ has been mapped to the unorthodox gauge-fixing
$A_\mu\,e^\mu_{\ \psi}=0$, so that $A_{t}\neq0$ in Minkowski space.
From $F$ we can read off the color-electric and the color-magnetic fields as $E_{a}:=F_{a0}$ and $B_{a}:=\frac{1}{2}\epsilon_{abc}F_{bc}$, respectively. They can be combined into a manageable form via the Riemann--Silberstein vector $\Vec{E}+\mathrm{i}\Vec{B}$, whose individual components read
\begin{equation}
\begin{aligned}
    (E{+}\im B)_{x}&\=-\frac{2(\im\varepsilon\Phi'+\Phi^2{-}1)}{(\lambda{-}2\im z)(\lambda{+}2\im z)^2}\,
    \Bigl\{ 2\bigl[ty{+}\im x(z{+}\im)\bigr] I_0 + 2\varepsilon\bigl[xy{+}\im t(z{+}\im)\bigr] I_1 + \varepsilon\bigl[t^2{-}x^2{+}y^2{+}(z{+}\im)^2\bigr] I_2 \Bigr\}\ ,\\
    (E{+}\im B)_{y}&\=\frac{2(\im\varepsilon\Phi'+\Phi^2{-}1)}{(\lambda{-}2\im z)(\lambda{+}2\im z)^2}\,
    \Bigl\{ 2\bigl[tx{-}\im y(z{+}\im)\bigr] I_0 + \varepsilon\bigl[t^2{+}x^2{-}y^2{+}(z{+}\im)^2\bigr] I_1 + 2\varepsilon\bigl[xy{-}\im t(z{+}\im)\bigr] I_2 \Bigr\}\ ,\\
    (E{+}\im B)_{z}&\=\frac{2(\im\varepsilon\Phi'+\Phi^2{-}1)}{(\lambda{-}2\im z)(\lambda{+}2\im z)^2}\,
    \Bigl\{ \im\bigl[t^2{+}x^2{+}y^2{-}(z{+}\im)^2\bigr] I_0 + 2\varepsilon\bigl[\im tx{-}y(z{+}\im)\bigr] I_1 + 2\varepsilon\bigl[\im ty{+}x(z{+}\im)\bigr] I_2 \Bigr\}\ ,
\end{aligned}
\end{equation}
where $\Phi=\Phi\bigl(\psi(t,x,y,z)\bigr)$ and $\lambda{\pm}2\im z=x^2{+}y^2{+}(z{\pm}\im)^2{-}t^2$.

\begin{figure}[ht!]
\centering
\includegraphics[width=7cm]{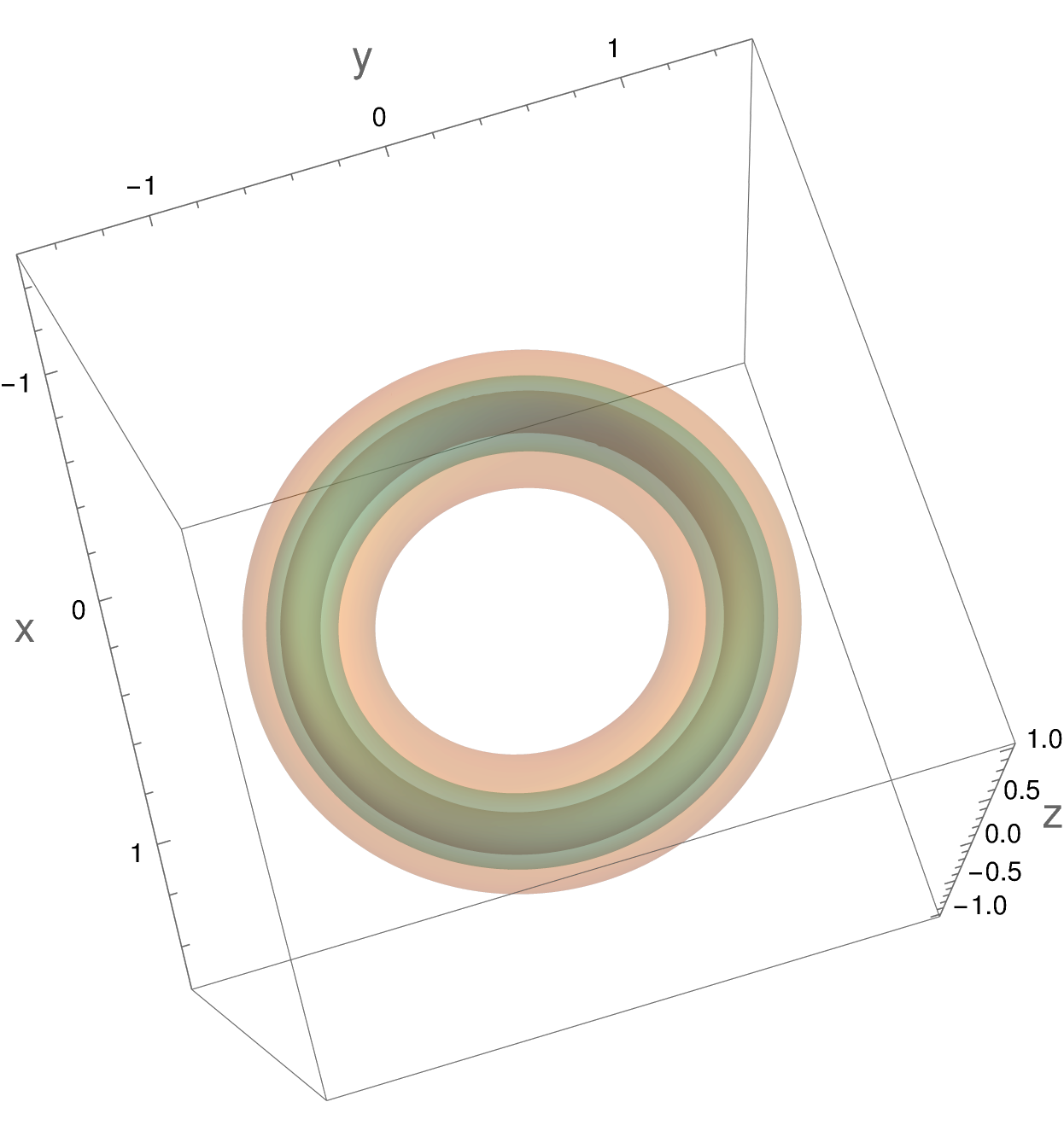}
\hfill
\includegraphics[width=7cm]{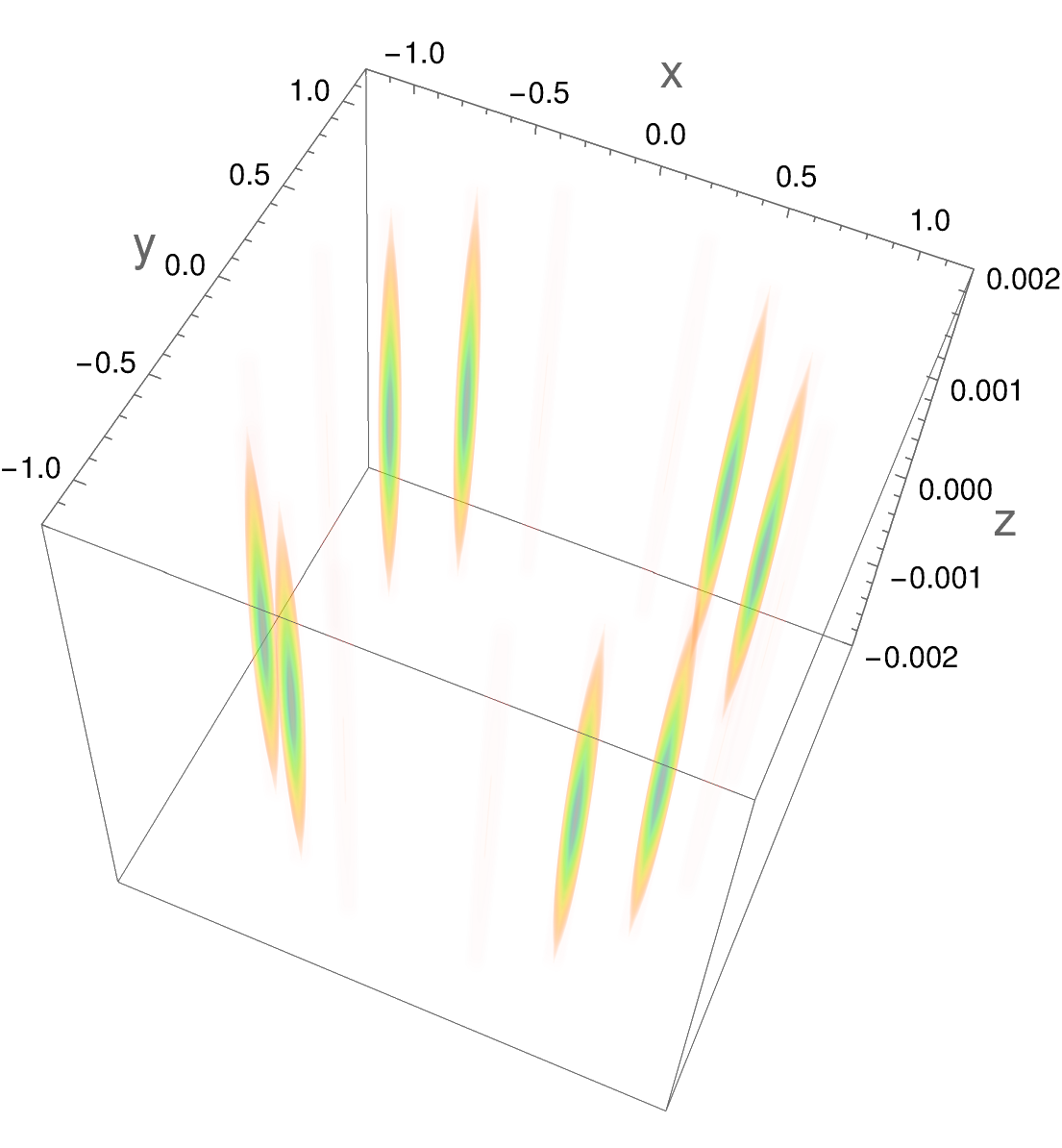}
\caption{Plots for the Minkowskian energy density proportional to $(\lambda^2{+}4z^2)^{-2}|_{t=0}$. Left: Level sets for the values $10$ (orange), $100$ (cyan), and $1000$ (brown). 
Right: Density plot emphasizing the maxima.}
\label{rhoNonAbel}
\end{figure}

\begin{figure}[ht!]
    \centering
    \includegraphics[width=5cm,height=5cm]{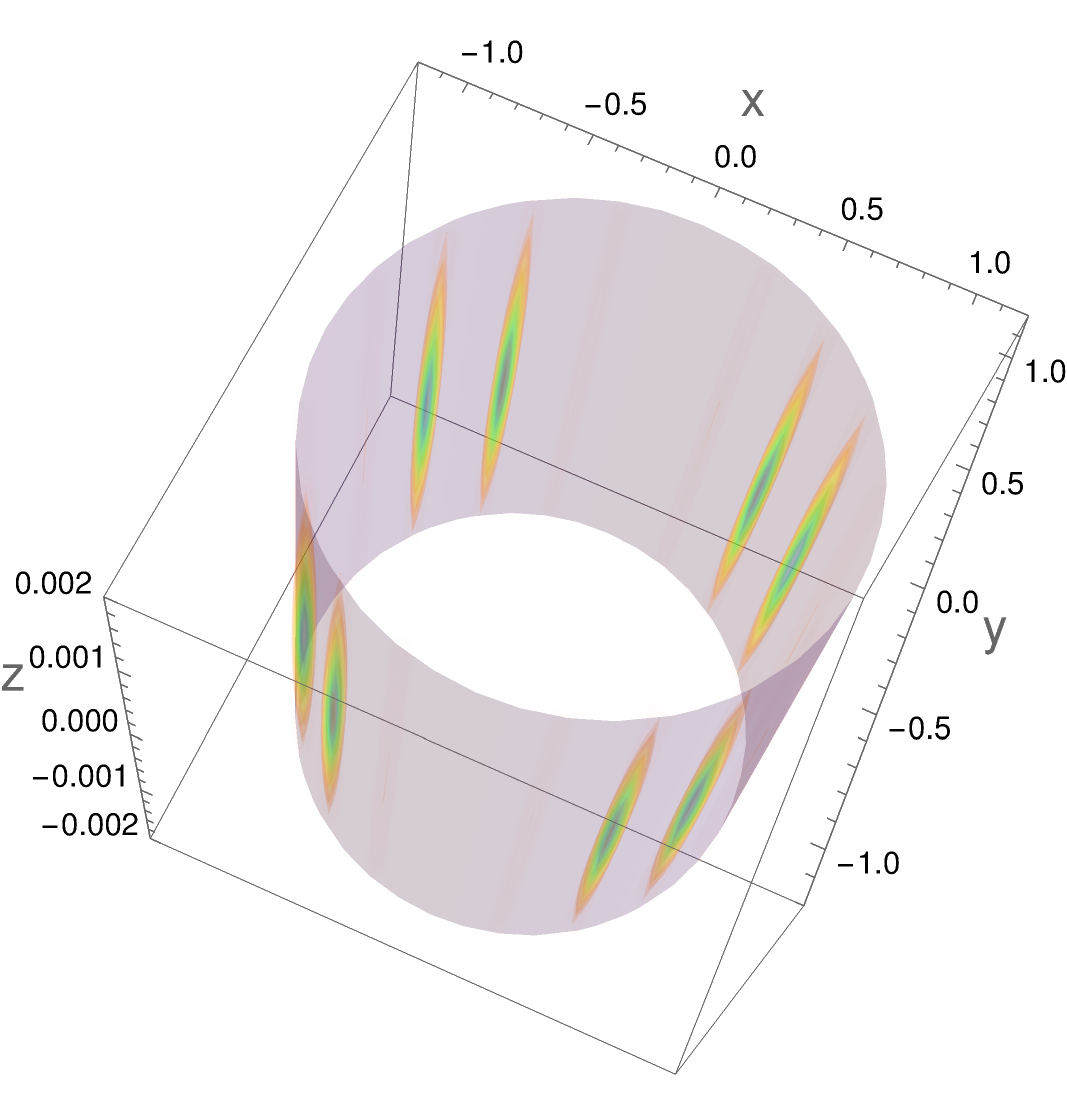}
    \quad
    \includegraphics[width=5cm]{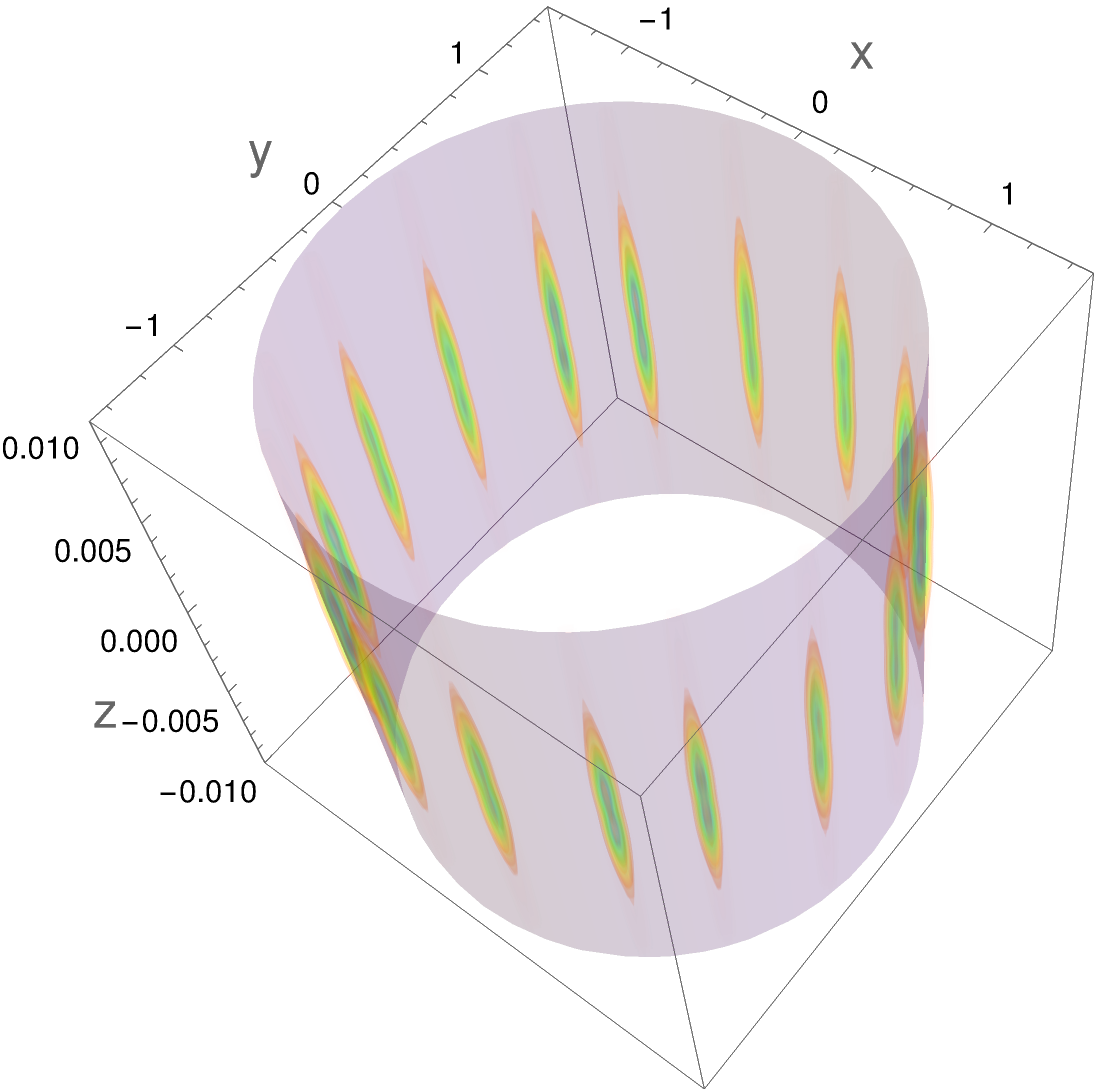}
    \quad
    \includegraphics[width=5cm,height=5cm]{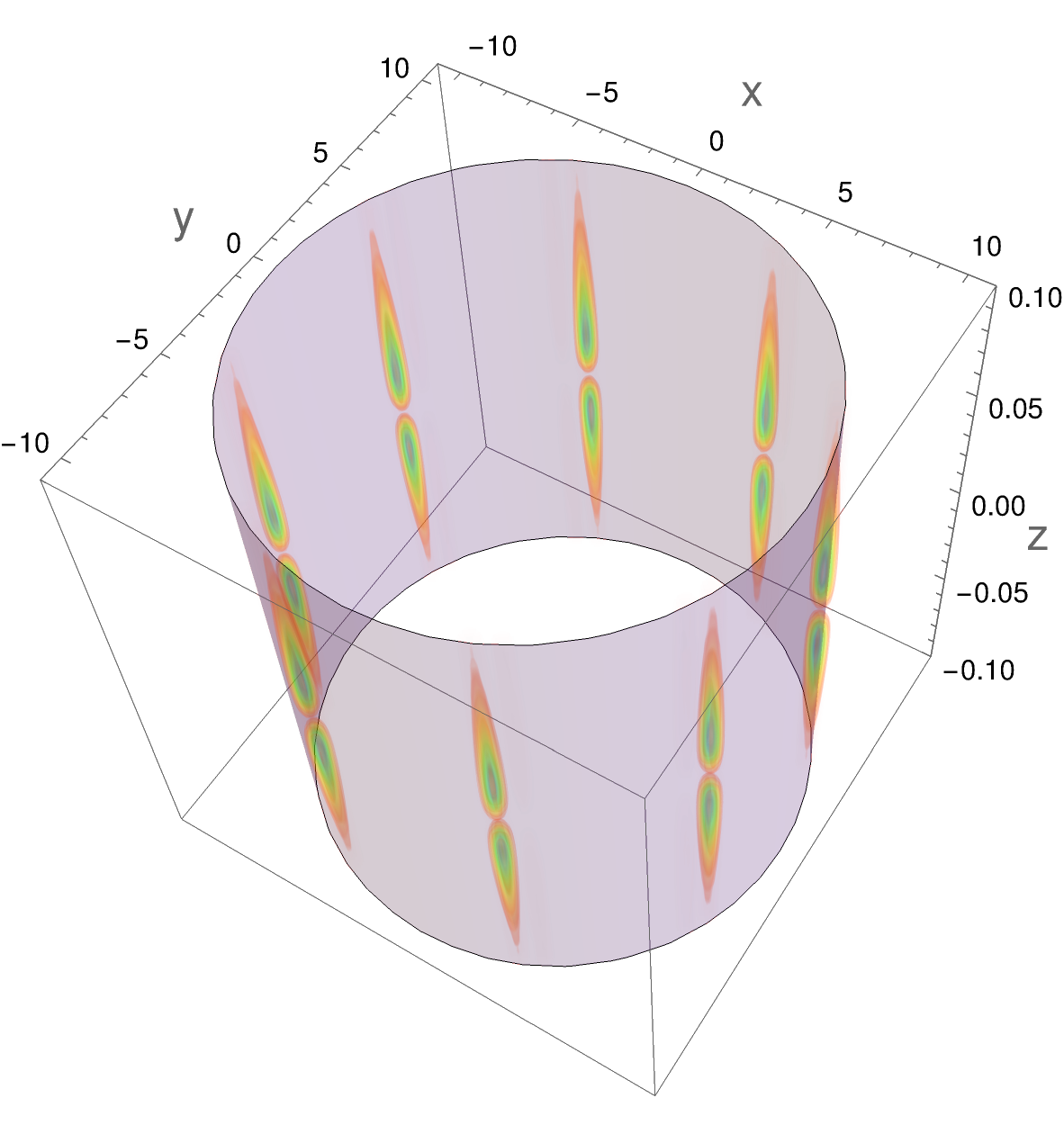}
    \caption{Demonstration of how the Minkowskian energy density is (largely) confined to the hypersurface $H^{1,2}_R$ (gray) at three distinct values of time: $t{=}0$ (left), $t{=}1$ (center) and $t{=}10$ (right).}
    \label{DenHyp}
\end{figure}

The expression for the energy-momentum tensor $T_{\mu\nu}$ can be obtained either directly via
\begin{align}\label{SEtensor}
    T_{\mu\nu}\=-\tfrac{1}{2g^{2}}\,\bigl( 
    \delta_\mu^{\ \rho}\delta_\nu^{\ \lambda}\eta^{\sigma\tau} -
    \tfrac14 \eta_{\mu\nu}\eta^{\rho\lambda}\eta^{\sigma\tau} \bigr)
    \,\mathrm{tr}\big(F_{\rho\sigma}F_{\lambda\tau}\bigr)
\end{align}
or by changing the basis in~\eqref{TAdS} while keeping track of the conformal factor arising when passing from the cylinder metric \eqref{metric1} to Minkowski metric \eqref{metric4}. 
The result does not depend on the value of~$\varepsilon$ and hence
is uniformly valid in the whole Minkowski space,
\begin{equation}
   \bigl(T_{\mu\nu}\bigr) \= \frac{8}{g^{2}}\,\frac{\rho}{(\lambda^2{+}4z^2)^3}
        \begin{pmatrix}
            \mathfrak{t}_{00} & -16\,t\,x\,z^2 & -16\,t\,y\,z^{2} & 8\,t\,z\,(\lambda{-}3z^{2}) \\[4pt]
            -16\,t\,x\,z^{2} & \mathfrak{t}_{11} & 16\,x\,y\,z^{2} & -8\,x\,z(\lambda{-}3z^{2}) \\[4pt] 
            -16\,t\,y\,z^{2} & 16\,x\,y\,z^{2} &  \mathfrak{t}_{22} & -8\,y\,z\,(\lambda{-}3z^{2}) \\[4pt] 
            8\,t\,z\,(\lambda{-}3z^{2}) &  -8\,x\,z(\lambda{-}3z^{2}) & -8\,y\,z\,(\lambda{-}3z^{2}) & \mathfrak{t}_{33}
        \end{pmatrix}\,,
\end{equation}
where $\rho(x)$ is given in~\eqref{TAdS}, $\lambda$ is in \eqref{1formsMink} and the diagonal components are
\begin{equation}
    \begin{split}
        \mathfrak{t}_{00}&\=\lambda^2 + 4z^2(1+4t^2)\ ,\\
        \mathfrak{t}_{22}&\=-\lambda^2-4z^2(1-4y^2)\ ,
    \end{split}
    \qquad
    \begin{split}
        \mathfrak{t}_{11}&\=-\lambda^2 -4z^2(1-4x^2)\ ,\\
    \mathfrak{t}_{33}&\=3\lambda^2-4z^2(1+4\lambda-4z^2)\ .
    \end{split}
\end{equation}
This also takes a nice compact form as
\begin{equation}
    \begin{split}
        \bigl(T_{\mu\nu}\bigr) \= \frac{8}{g^{2}}\,\frac{\rho}{(\lambda^2{+}4z^2)^3}
        \begin{pmatrix}
            \mathfrak{t}_{\alpha\beta} & \mathfrak{t}_{\alpha 3} \\
        \mathfrak{t}_{3\alpha} & \mathfrak{t}_{33}
        \end{pmatrix}
    \end{split}
    \quad\with \biggl\{ \
    \begin{split}
        \mathfrak{t}_{\alpha\beta} &\= -\eta_{\alpha\beta}(\lambda^2{+}4z^2)+16x_\alpha x_\beta z^2\,,\\
        \mathfrak{t}_{3\alpha} &\= \mathfrak{t}_{\alpha 3} \= -8x_\alpha z\,(\lambda{-}3z^2)\ .
    \end{split}
\end{equation}
Clearly, the energy-momentum tensor is singular at the intersection of the hypersurfaces $z{=}0$ and $\lambda{=}0$, which is the two-dimensional Lorentzian hyperboloid 
\begin{equation} \label{singular}
    \bigl(t,x,y,z\bigr) \= \bigl(t,\,\sqrt{1{+}\smash{t^2}}\,\cos\phi,\,\sqrt{1{+}\smash{t^2}}\,\sin\phi,\,0\bigr)\ .
\end{equation}
In Figure~\ref{rhoNonAbel} we have plotted the Minkowskian energy density at $t{=}0$, which is simply $(\lambda^2{+}4z^2)^{-2}|_{t=0}$ (up to a factor),  and three level sets of it, while in Figure~\ref{DenHyp} we show that it remains mostly confined to the boundary hyperbola $H^{1,2}_R$ at all times. Furthermore, the conserved field energy $E$ can be computed by integrating $2g^2\,T_{00}$ over any spatial slice, most suitably the $t{=}\tau{=}0$ slice.
The result is clearly divergent due to the singularity~\eqref{singular},
\begin{equation}
    \begin{aligned}
        E &\= 2g^2\int\mathrm{d}^{3}x\ T_{00}|_{t{=}0}
        \=-\tfrac{1}{2} \int_{\mathbb{R}^{3}}\!\!\!\mathrm{d}^{3}x\ \mathrm{tr}(E_{a}E_{a}{+}B_{a}B_{a})|_{t{=}0}\\
        &\= \rho \int_{S^{3}}\!\!\!\mathrm{d}^{3}\Omega\ \frac{1-\cos\chi}{(1{-}\sin^{2}\!\chi\sin^{2}\!\theta)^{2}}
        \= 2\pi\,\rho\int^{\pi}_{0}\!\!\mathrm{d}\chi\int^{\pi}_{0}\!\!\mathrm{d}\theta\ \frac{\sin^{2}\!{\chi}\,\sin{\theta}\;(1-\cos\chi)}{(1-\sin^{2}\!\chi\,\sin^{2}\theta)^{2}}\ .
    \end{aligned}
\end{equation}

\begin{figure}[H]
\centering
\includegraphics[width=8cm,height=8cm]{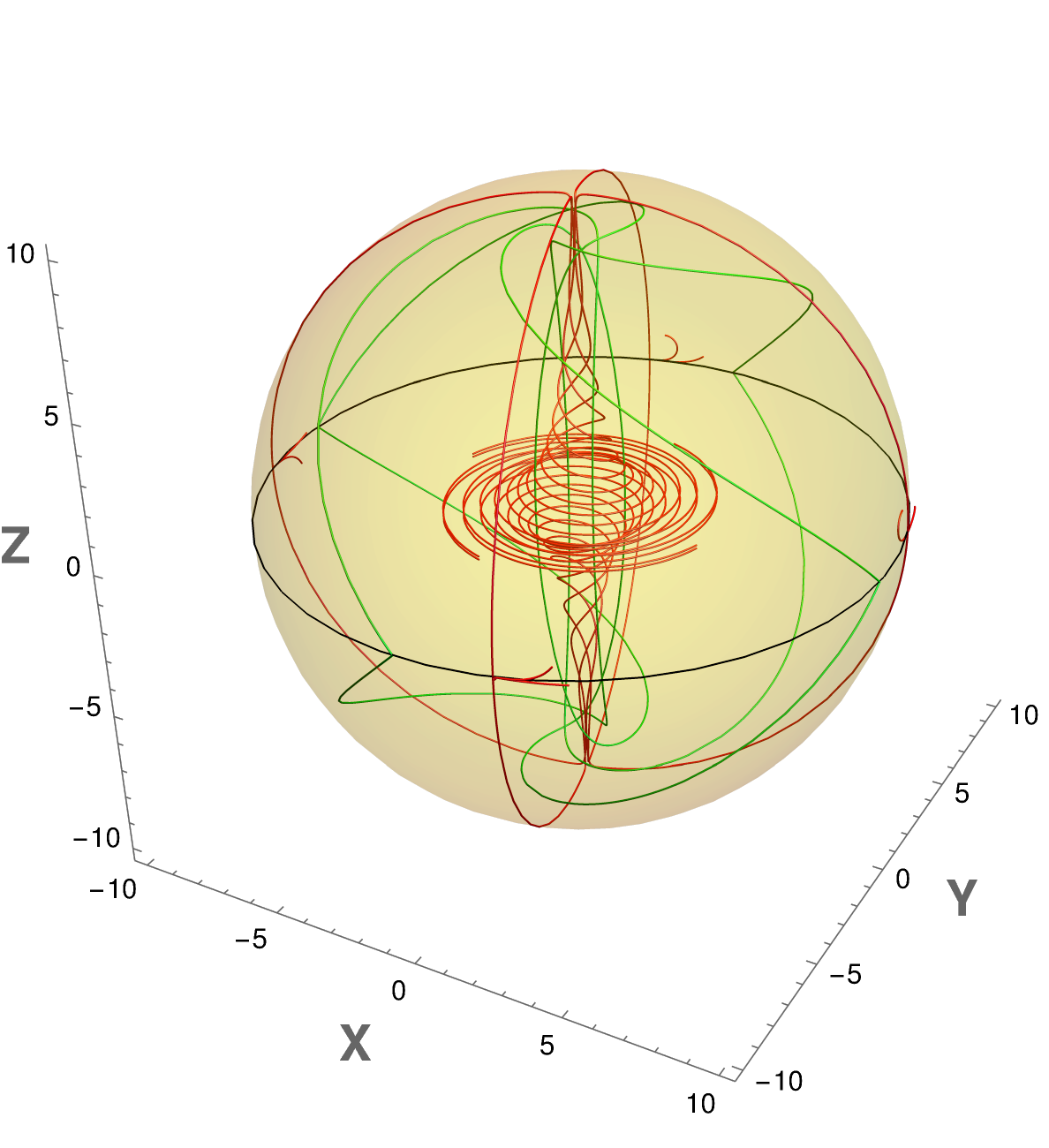}
\caption{Typical electric (red) and magnetic (green) field lines for the Abelian solution \eqref{AbelRS} with $t{=}10$ and $\psi_0{=}\sfrac{\pi}{2}$ inside the boundary sphere $r{=}\sqrt{101}$. The fields diverge at the singular equatorial circle (black), which is the intersection of the boundary sphere with $z{=}0$ plane.}
\label{fieldLines}
\end{figure}

\newpage
\noindent
\textbf{Abelian solutions.} As before, the gauge potential and the field strength can be expressed in the Minkowski basis~$\{\diff x^\mu\}$ easily using the transformation factors $e^{\alpha}_{\ \mu}$ and $e^{\psi}_{\ \mu}$ computed earlier. 
Without loss of generality we set $f{=}1$ in~\eqref{abelianAF} and conveniently place~$\varepsilon$ to obtain
\begin{equation} \label{tildeF}
\begin{aligned}
    \widetilde{A}&\=-\tfrac12\,\cos2\bigl(\psi(x){+}\varepsilon\psi_{0}\bigr)\,e^{0}_{\ \mu}\;\mathrm{d}x^{\mu}
    \qquad\mathrm{and}\\[4pt]
    \widetilde{F}&\= \bigl\{ \sin2\bigl(\psi(x){+}\varepsilon\psi_{0}\bigr)\,e^{\psi}_{\ \mu} e^0_{\ \nu}
    -\cos2\bigl(\psi(x){+}\varepsilon\psi_{0}\bigr)\,e^{1}_{\ \mu}e^{2}_{\ \nu}\big\}\;\mathrm{d}x^{\mu}\wedge \mathrm{d}x^{\nu}\ .
\end{aligned}
\end{equation}
Employing the relation~\eqref{AdStoMink} we evaluate
\begin{equation}
    \begin{aligned}
    \cos2\bigl(\psi(x){+}\varepsilon\psi_{0}\bigr)
    \=\frac{1}{\lambda^2{+}4z^2}\,\big(4z\,\lambda\sin{2\psi_{0}}+(\lambda^{2}{-}4z^{2})\cos{2\psi_{0}}\big)
    \ =:\ h_1(x)\ ,\\
    \sin2\bigl(\psi(x){+}\varepsilon\psi_{0}\bigr)
    \=\frac{-\varepsilon}{\lambda^2{+}4z^2}\,\big(4z\,\lambda\cos{2\psi_{0}}-(\lambda^{2}{-}4z^{2})\sin{2\psi_{0}}\big)
    \ =:\ h_2(x)\ ,
\end{aligned}
\end{equation}
which are regular throughout Minkowski space.
We can then easily read off electric $\widetilde{E}_{a}{:=}\widetilde{F}_{a0}$ and magnetic $\widetilde{B}_{a}{:=}\tfrac{1}{2}\epsilon_{abc}\widetilde{F}_{bc}$ fields from \eqref{tildeF} to arrive at
\begin{equation}
\begin{split}
    \widetilde{E}_{1}&\=h_{1}(x)\,q_{1}(x)+\varepsilon\,h_{2}(x)\,p_{1}(x)\ ,\\
    \widetilde{E}_{2}&\=h_{1}(x)\,q_{2}(x)+\varepsilon\,h_{2}(x)\,p_{2}(x)\ ,\\
    \widetilde{E}_{3}&\=h_{1}(x)\,q_{3}(x)+\varepsilon\,h_{2}(x)\,p_{3}(x)\ ,
\end{split}
\qquad
\begin{split}
    \widetilde{B}_{1}&\=-h_{1}(x)\,p_{1}(x)+\varepsilon\,h_{2}(x)\,q_{1}(x)\ ,\\
    \widetilde{B}_{2}&\=-h_{1}(x)\,p_{2}(x)+\varepsilon\,h_{2}(x)\,q_{2}(x)\ ,\\
    \widetilde{B}_{3}&\=-h_{1}(x)\,p_{3}(x)+\varepsilon\,h_{2}(x)\,q_{3}(x)\ .
\end{split}
\end{equation}
where the functions $p_1(x),\ldots,q_3(x)$ are given by
\begin{equation}
    \begin{split}
        q_{1}(x)&\=\sfrac{8}{(\lambda^2+4z^2)^2}\big(2xz^2-\lambda(x{-}ty)\big)\ ,\\ q_{2}(x)&\=\sfrac{8}{(\lambda^2+4z^2)^2}\big(2yz^2-\lambda(y{+}tx)\big)\ ,\\
        q_{3}(x)&\=\sfrac{-16}{(\lambda^2+4z^2)^2}\big(z\,(x^{2}{+}y^{2})\big)\ ,
    \end{split}
    \qquad\begin{split}
        p_{1}(x)&\=\sfrac{-8}{(\lambda^2+4z^2)^2}\big(\lambda xz +2z(x{-}ty)\big)\ ,\\
        p_{2}(x)&\=\sfrac{-8}{(\lambda^2+4z^2)^2}\big(\lambda yz+2z(y{+}tx)\big)\ ,\\
        p_{3}(x)&\=\sfrac{-4}{(\lambda^2+4z^2)^2}\big(\lambda^2-2\lambda(x^2{+}y^2)+4z^2\big)\ .\\
    \end{split}
\end{equation}
The structure of these fields reflects the presence of electromagnetic duality: 
$\widetilde{E}_{a}\rightarrow\widetilde{B}_{a}\rightarrow-\widetilde{E}_{a}$ 
via $h_{1}\rightarrow\varepsilon h_{2}\rightarrow-h_{1}$ and $h_{2}\rightarrow-\varepsilon h_{1}\rightarrow-h_{2}$,
which in the orthonormal frame means 
$\widetilde{\mathcal{E}}_{0}\rightarrow\widetilde{\mathcal{B}}_{0}\rightarrow-\widetilde{\mathcal{E}}_{0}$. 
As before, the Riemann--Silberstein vector $\vec{S}$ takes a simple form,
\begin{equation}\label{AbelRS}
    \vec{S}\ :=\ \Vec{\widetilde{E}}+\im\Vec{\widetilde{B}} \= \frac{4\ep^{2\im\psi_0}}{(\lambda{+}2\im z)^3}
    \begin{pmatrix}
        -2\,\bigl(t\,y+\im x\,(z{+}\im)\bigr)\\[4pt]
        2\,\bigl(t\,x-\im y\,(z{+}\im)\bigr)\\[4pt]
        \im\,\bigl(t^2+x^2+y^2-(z{+}\im)^2\bigr)
        \end{pmatrix}\ ,
\end{equation}
and is also singular at the hyperboloid~\eqref{singular}. This is clearly demonstrated with field lines in Figure \ref{fieldLines}.

Let us consider a special case of this field configuration, namely $\psi_0=z=0$,
\begin{equation}\label{AbelAdS}
    \frac{4}{(x^2{+}y^2{-}t^2{-}1)^3}\begin{pmatrix}
        2(x-t\,y) \\[4pt] 2(y+t\,x) \\[4pt] t^2{+}x^2{+}y^2{+}1
    \end{pmatrix}\ ,
\end{equation}
consisting of two electric and one magnetic (the last one) fields in $2{+}1$~dimensions. We can consider these as three gauge fields on AdS$_3$, equipped with coordinates $(x^0,x^1,x^2)$. Interestingly, the identification
\begin{equation}
    x^0 \= -t\ ,\quad x^1 \= -y \und x^2 \= x
\end{equation}
reproduces the magnetic field of a recently constructed magnetic vortex~\cite[eq. (5.15)]{RS18}. The latter arises from a pair $(\Psi,A)$ of spinor and gauge fields satisfying a Dirac and a non-linear equation respectively on SU$(1,1)$, that is then pulled back to Minkowski space $\R^{1,2}$ via inverse stereographic projection from $\textrm{AdS}_3\cong\textrm{SU}(1,1)/\Z_2$.\footnote{The Authors of \cite{RS18} employ a ``mostly-minus'' Minkowski signature and also set the AdS scale to unity, $R=1$. They interpret the components in \eqref{AbelAdS} as magnetic fields arising from $F\sim e^1\we e^2$.} It is worth noting that their Abelian field is defined only for $\lambda_{z=0}<0$, i.e.~inside the hyperbola embedded in $\R^{1,2}$, as is also clear from \cite[Figure 4]{RS18}. Our Abelian solution \eqref{AbelAdS}, on the other hand, is valid on the full Minkowski space $\R^{1,2}$ except on the hyperbola $\lambda_{z=0}=0$. 

\begin{figure}[H]
\centering
\includegraphics[width=5cm,height=5cm]{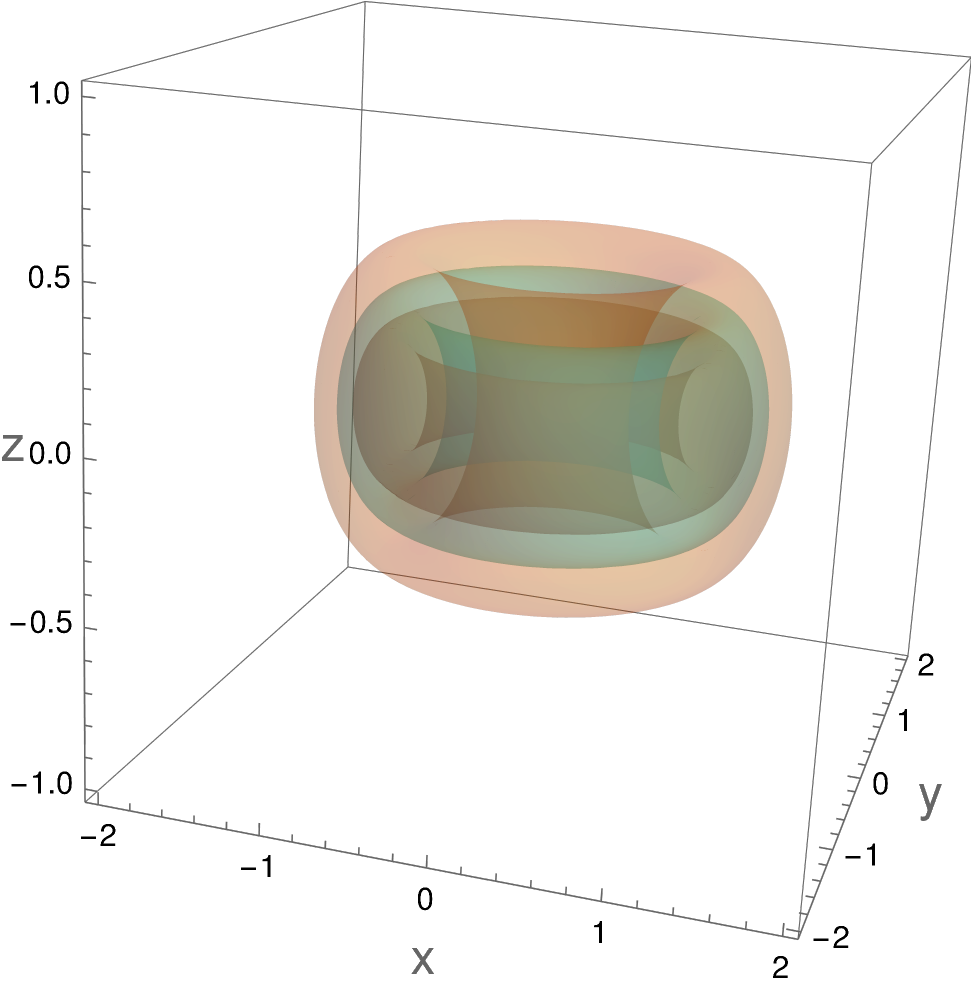}
\qquad\qquad\qquad
\includegraphics[width=5cm,height=5cm]{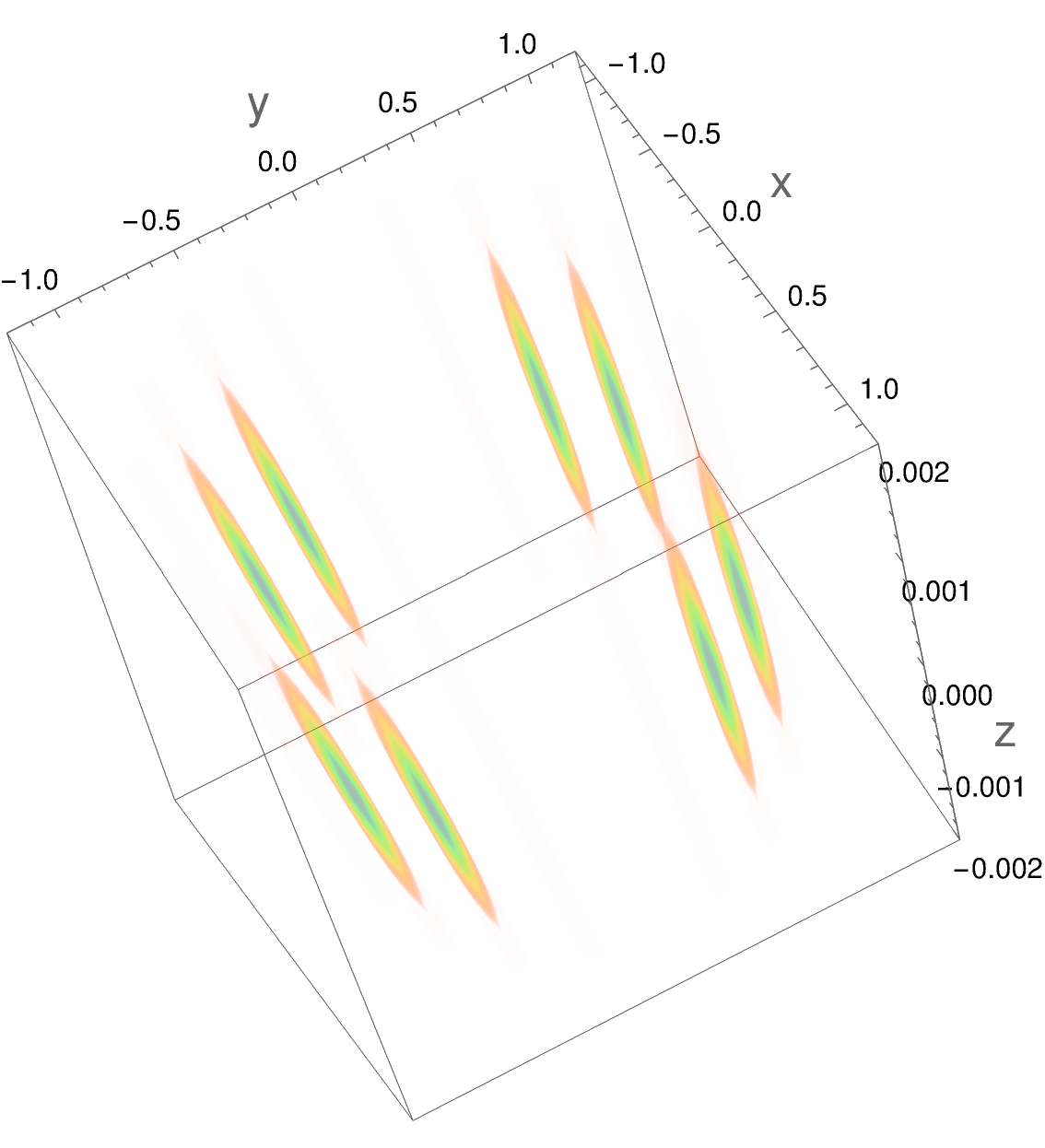}
\caption{Minkowskian energy density $\tfrac{1}{2}\Vec{S}^{\dagger}\Vec{S}(t{=}0)$. Left: Level sets for the values $10$ (orange), $100$ (cyan), and $1000$ (brown). Right: Density plot emphasizing the maxima.}
\label{rhoAbel}
\end{figure}

The Minkowskian energy density is readily computed from \eqref{AbelRS} via
\begin{equation}
\begin{aligned}
    \tfrac{1}{2}\Vec{S}^{\dagger}\Vec{S}
    &\=\tfrac{1}{2}(\widetilde{E}_{a}\widetilde{E}_{a}+\widetilde{B}_{a}\widetilde{B}_{a}) \= \tfrac{8}{(\lambda^2{+}4z^2)^3}\big(\lambda^2+8(1{+}t^2)(x^2{+}y^2)+4z^2\big)\ .
\end{aligned}
\end{equation}
We also provide below a density plot and level sets for this energy density in Figure~\ref{rhoAbel}.
\noindent
The divergent expression for the total energy then becomes 
\begin{align}
    \widetilde{E}\=\tfrac{1}{2}\int\!\mathrm{d}^{3}x\ \Vec{S}^{\dagger}\Vec{S}|_{t{=}0}
    \=\tfrac{1}{2}\int\!\mathrm{d}^{3}\Omega\ (1-\cos{\chi})\,\frac{1+\sin^{2}\!{\chi}\sin^{2}\!{\theta}}{(1-\sin^{2}\!{\chi}\sin^{2}\!{\theta})^{3}}\ .
\end{align}
Unlike the non-Abelian case, here we get a rather bulky energy-momentum tensor $T_{\mu\nu}$ \eqref{SEtensor} whose (independent) components read as follows,
\begin{equation}
    \begin{split}
        T_{00}&=\sfrac{4}{g^2(\lambda^2+4z^2)^3}\big(\lambda^2+8(1{+}t^2)(x^2{+}y^2)+4z^2\big)\,,\\
        T_{02}&=\sfrac{16}{g^2(\lambda^2+4z^2)^3}\big((x{-}ty)(\lambda{+}2{+}2t^2)+2tyz^2\big)\,,\\
        T_{11}&=\sfrac{4}{g^2(\lambda^2+4z^2)^3}\big(\lambda^2+8x^2(t^2{-}z^2)+8y(y{+}2tx)+4z^2\big)\,,\\
        T_{22}&=\sfrac{4}{g^2(\lambda^2+4z^2)^3}\big(\lambda^2+8y^2(t^2{-}z^2)+8x(x{-}2ty)+4z^2\big)\,,\\
        T_{23}&=\sfrac{16}{g^2(\lambda^2+4z^2)^3}\big(yz(\lambda{-}2z^2)-2tz(x{-}ty)\big)\,,\\
    \end{split}
    \quad
    \begin{split}
        T_{01}&=\sfrac{-16}{g^2(\lambda^2+4z^2)^3}\big((tx{+}y)(\lambda{+}2{+}2t^2)-2txz^2\big)\,,\\
        T_{12}&=\sfrac{32}{g^2(\lambda^2+4z^2)^3}\big({-}xy(z^2{-}t^2{+}1)+t(y^2{-}x^2)\big)\,,\\
        T_{03}&=\sfrac{-32}{g^2(\lambda^2+4z^2)^3}\,tz(x^{2}{+}y^{2})\,,\\
        T_{13}&=\sfrac{16}{g^2(\lambda^2+4z^2)^3}\big(xz(\lambda{-}2z^2)+2tz(y{+}tx)\big)\,,\\
        T_{33}&=\sfrac{-4}{g^2(\lambda^2+4z^2)^3}\big(\lambda^2-4z^2(2x^2{+}2y^2{-}1)\big)\ .
    \end{split}
\end{equation}
Clearly, the stress-energy tensor is also divergent here on the hyperboloid \eqref{singular}. We conclude this section with a remark on the singularity of the field expressions and their stress-energy tensors in both the Abelian and non-Abelian cases: although these are singular at the intersection $\{\lambda{=}0\}\cap\{z{=}0\}$ this singularity, nevertheless, is rather mild when compared with the $\lambda{=}0$ singularity that we started with, e.g.~in \eqref{metric4}.

\bigskip
\noindent {\bf Acknowledgements}

\noindent
This work was supported by the Deutsche Forschungsgemeinschaft grant LE 838/19.
KK is grateful to the Institute for Theoretical Physics at Leibniz University for support during the tenure of this project. SH thanks Mahir Ertürk for useful discussions. KK thanks Calum Ross for pointing out his related works and several insightful discussions.

%~~~~~~~~~~~~~~~~~~~~~~~~~~~~~~~~~~~~~~~~~~~~~~~~~~~

%~~~~~~~~~~~~~~~~~~~~~~~~~~~~~~~~~~~~~~~~~~~~~~~~~~~
\end{document}